\documentclass[longauth]{aa}
\usepackage{orcidlink}
\usepackage{graphicx}
\usepackage{pdflscape}
\usepackage{txfonts}
\usepackage{amsmath}  
\usepackage{natbib}
\usepackage{booktabs}
\usepackage{longtable}
\usepackage{array}
\usepackage{hyperref}

\hypersetup{
    colorlinks=true,
    linkcolor=blue,
    citecolor=blue,
    filecolor=magenta,
    urlcolor=cyan,
    pdftitle={Ellipsoidal modulation and multi-wavelength activity in the pre-cataclysmic binary RX~J1553.0+4457},
}

\newcommand{\distpc}{36.833}              % distance (pc), fixed in the SED fit
\newcommand{\av}{0.03}                    % fixed foreground extinction

\newcommand{\wdteff}{7000}                % WD effective temperature (K), adopted constrained fit
\newcommand{\wdlogg}{8.0}                 % fixed WD log g in the adopted SED solution

\newcommand{\mdteff}{3200}                % M-dwarf effective temperature (K)
\newcommand{\mdlogg}{4.5}                 % M-dwarf log g

\newcommand{\vgfb}{3.26}                  % Vgfb value for the adopted two-component fit

\newcommand{\EP}{\textit{Einstein Probe}}

\begin{document}

  \title{Ellipsoidal modulation and multi-wavelength activity in the pre-cataclysmic binary RX~J1553.0+4457}

\titlerunning{Ellipsoidal modulation and activity in RX~J1553.0+4457}
\authorrunning{Wu et al.}

\author{
S.-Y. Wu\inst{1,2}\thanks{\normalfont Corresponding author. Email: wusiyu.11@outlook.com}\,\orcidlink{0009-0004-7113-8258}
\and
M. Gritsevich\inst{1,3,4}\orcidlink{0000-0003-4268-6277}
\and
Q.-H. Lao\inst{5}
\and
A. J. Castro-Tirado\inst{1,6}
\and
Z. Li\inst{7}\,\orcidlink{0000-0003-0418-8461}
\and
Y.-D. Hu\inst{8}\thanks{\normalfont Corresponding author. Email: huyoudong072@hotmail.com}
\and
I.~P\'{e}rez-Garc\'{\i}a\inst{1,6}
\and
R. S\'{a}nchez-Ram\'{\i}rez\inst{1}
\and
N. Castro-Segura\inst{9}
\and
E.~J.~Fern\'{a}ndez-Garc\'{\i}a\inst{1}
\and
M.~D.~Caballero-Garc\'{\i}a\inst{1}
\and
S. Guziy\inst{1}
\and
I. Olivares\inst{1}
\and
J. D. Sakowska\inst{1}
\and
G. Garc\'{\i}a-Segura\inst{10}
\and
D. Hiriart\inst{10}
\and
W. H. Lee\inst{11}\,\orcidlink{0000-0002-2467-5673}
\and
P. J. Meintjes\inst{12}
\and
H.~J.~van Heerden\inst{12}\,\orcidlink{0009-0004-9747-7215}
\and
A. Mart\'{i}n-Carrillo\inst{13}
\and
L. Hanlon\inst{13}
\and
A. Maury\inst{14}
\and 
L. Hern\'{a}ndez-Garc\'{\i}a\inst{15,16}
\and
I. M. Carrasco-Garc\'{\i}a\inst{17}
\and
S. Castillo-Carri\'{o}n\inst{18}\,\orcidlink{0000-0002-0258-2046}
\and
A. Castell\'{o}n\inst{19}
\and
S. B. Pandey\inst{20}
\and
C. J. P\'{e}rez del Pulgar\inst{6}
\and
A. J. Reina\inst{6}
\and
J.-M. Bai\inst{7}
\and
Y.-F. Fan\inst{7}
\and
C.-J. Wang\inst{7}
\and
Y.-X. Xin\inst{7}
\and
D.-R. Xiong\inst{7}
\and
X.-H. Zhao\inst{7}
\and
J. Mao\inst{7}
\and
B.-L. Lun\inst{7}
\and
K. Ye\inst{7}
\and
C.-Z. Cui\inst{5}
\and
A. F. Valeev\inst{21}
\and
B.-B. Zhang\inst{22}
\and
T.-R. Sun\inst{23}\,\orcidlink{0000-0003-1166-3814}
}

\institute{
Instituto de Astrof\'{\i}sica de Andaluc\'{\i}a, Consejo Superior de Investigaciones Cient\'{\i}ficas (IAA-CSIC), Glorieta de la Astronom\'{\i}a, s/n, 18080 Granada, Spain
\and
Department of Physics and Mathematics, Facultad de Ciencias, Universidad de Granada, Avda. de Fuentenueva s/n, 18071 Granada, Spain
\and
Faculty of Science, University of Helsinki, Gustaf Hallstr\"{o}min katu 2, FI-00014 Helsinki, Finland
\and
Institute of Physics and Technology, Ural Federal University, Mira str. 19, 620002 Ekaterinburg, Russia
\and
National Astronomical Observatories, Chinese Academy of Sciences, 20A Datun Road, Beijing 100101, People's Republic of China
\and
Departamento de Ingenier\'{\i}a de Sistemas y Autom\'{a}tica, Unidad Asociada al CSIC por el IAA, Escuela de Ingenier\'{\i}as Industriales, Universidad de M\'{a}laga, calle Dr. Ortiz Ramos s/n, 29071 M\'{a}laga, Spain
\and
Yunnan Observatories, Chinese Academy of Sciences, 396 Yangfangwang, Guandu District, Kunming 650216, People's Republic of China
\and
Guangxi Key Laboratory for Relativistic Astrophysics, School of Physical Science and Technology, Guangxi University, Nanning 530004, China
\and
Department of Physics, University of Warwick, Coventry, CV4 7AL, United Kingdom
\and
Instituto de Astronom\'{\i}a, Universidad Nacional Aut\'{o}noma de M\'{e}xico (IA-UNAM), carretera Tijuana-Ensenada, km. 107, C.P. 22860 Ensenada, Baja California, M\'{e}xico
\and
Instituto de Astronom\'{\i}a, Universidad Nacional Aut\'{o}noma de M\'{e}xico (IA-UNAM), Apartado Postal 70-264, CDMX, C.P. 04510 M\'{e}xico DF, M\'{e}xico
\and
Department of Physics, University of the Free State, Nelson Mandela Drive, 9301 Bloemfontein, South Africa
\and
University College Dublin, School of Physics, L.M.I. Main Building, Beech Hill Road, Dublin 4, D04 P7W1 Dublin, Ireland
\and
San Pedro de Atacama Celestial Explorations, 2RXC+23, Solor, San Pedro de Atacama, Antofagasta, Chile
\and
Instituto de Estudios Astrof\'{\i}sicos, Facultad de Ingenier\'{\i}a y Ciencias, Universidad Diego Portales, Av. Ej\'{e}rcito Libertador 441, Santiago, Chile
\and
Centro Interdisciplinario de Data Science, Facultad de Ingenier\'{\i}a y Ciencias, Universidad Diego Portales, Av. Ej\'{e}rcito Libertador 441, Santiago, Chile
\and
Universidad Alfonso X El Sabio, UAX Mare Nostrum, Camino de la T\'{e}rmica 90, 29004 M\'{a}laga, Spain
\and
Servicio Central de Inform\'{a}tica, Aulario L\'{o}pez de Pe\~{n}alver, calle Jim\'{e}nez Fraud 10, Campus de Teatinos, Universidad de M\'{a}laga, 29071 M\'{a}laga, Spain
\and
Departamento de \'{A}lgebra, Geometr\'{\i}a y Topolog\'{\i}a, Universidad de M\'{a}laga, Bulevar Louis Pasteur 31, Campus de Teatinos, 29071 M\'{a}laga, Spain
\and
Aryabhatta Research Institute of Observational Sciences (ARIES), Manora Peak, Nainital, Uttarakhand 263001, India
\and
Special Astrophysical Observatory, Russian Academy of Sciences, Nizhnii Arkhyz, Karachay-Cherkessia, 369167 Russia
\and
School of Physics, Nanjing University, 22 Hankou Road, Nanjing 210023, China
\and
Purple Mountain Observatory, Chinese Academy of Sciences, Nanjing 210023, China
}

   \date{}

% \abstract{}{}{}{}{} 
% 5 {} token are mandatory
 
\abstract
{RX~J1553.0+4457 (TMTS~J15530469+4457458) is a detached post-common-envelope binary with a cool white dwarf and an active late-type companion. Its orbital parameters are known, but the origin of its broadband optical modulation, short-timescale activity, and ultraviolet-to-mid-infrared spectral energy distribution (SED) remains unclear.}
{We test whether the dominant long-baseline \textit{TESS} waveform is ellipsoidal, characterize the minute-to-hour optical activity, and examine whether the SED requires any non-stellar component.}
{We combine high-cadence BOOTES multi-band photometry, six sectors of public \textit{TESS} full-frame imaging, \textit{Einstein Probe}/FXT X-ray data, single-epoch CAFOS spectroscopy, and archival photometry. We model the BOOTES flare decays, analyse the \textit{TESS} timing and flare statistics with injection--recovery tests, fit time-resolved X-ray spectra, and model the SED with white-dwarf plus M-dwarf photospheres.}
{The BOOTES data reveal two short optical flares separated by about 3~h, with amplitudes of roughly 1--1.5~mag and faster decay at shorter wavelengths. The combined \textit{TESS} light curve shows a stable signal at $P=0.083782$~d, the first harmonic of the known orbital period. Its dominance at $2f_{\rm orb}$ and double-wave morphology indicate ellipsoidal modulation from a tidally distorted companion. The \textit{TESS} flare energies are consistent with active M dwarfs. The \textit{Einstein Probe}/FXT spectra show a factor of ${\sim}4$ decline in the 0.3--10\,keV flux, mainly through decreasing emission measures. The SED matches the cool-white-dwarf plus late-type-companion configuration and shows no hot-continuum or mid-infrared excess.}
{RX~J1553.0+4457 is best interpreted as a near-Roche-lobe-filling pre-cataclysmic binary whose stable optical waveform is dominated by ellipsoidal modulation, while its short-timescale flares are mainly magnetic activity on the late-type companion. The data do not require a luminous accretion disc or irradiation-powered optical component, although weak wind-fed or intermittent accretion cannot be excluded.}


\maketitle

%
%-------------------------------------------------------------------

\section{Introduction}
\label{sec:intro}

Detached binaries consisting of a white dwarf (WD) and an M dwarf, many of which are post-common-envelope binaries (PCEBs), provide an important laboratory for studying close-binary evolution, angular-momentum loss, magnetic activity in tidally influenced late-type stars, and the onset of weak accretion in systems close to, but not necessarily in, stable mass transfer \citep{RebassaMansergas2007,RebassaMansergas2010,Zorotovic2010}. High-cadence photometry is particularly useful in such systems because it can separate coherent orbital modulation from stochastic activity such as flares and flickering-like variability.

RX~J1553.0+4457 (TMTS\,J15530469+4457458) was first identified as an X-ray source in the \textit{ROSAT} All-Sky Survey \citep{Voges1999} and has recently been studied in detail by \citet{Liu2025}. They showed that the system is a detached PCEB containing a $\sim 0.56\,M_\odot$ white dwarf and an M4 companion, and derived an orbital period of $P_{\rm orb}=0.16756456\pm0.00000006$~d. In the same study, phase-resolved spectroscopy revealed double-peaked Balmer emission, which was interpreted as evidence for wind-driven accretion rather than for a fully developed accretion disc. This interpretation has clear precedents in detached PCEBs showing multiple Balmer-emission components and weak wind-accretion signatures \citep{Tappert2011,Ribeiro2013}.

Despite these strong constraints on the basic system architecture, several observational aspects remain independent of the orbital solution and have not yet been characterized in a uniform way. These include the minute-to-hour optical activity, the colour evolution of individual flares, the long-baseline stability of the dominant modulation seen with the \textit{Transiting Exoplanet Survey Satellite} (\textit{TESS}), the short-term behaviour of the X-ray-emitting plasma during the \textit{Einstein Probe} observation, and the question of whether the ultraviolet-to-mid-infrared spectral energy distribution (SED) requires any component beyond the two stellar photospheres. These measurements probe the current activity state of the system, rather than its already established basic orbital architecture.

Observationally, distinguishing magnetic activity from weak accretion in detached WD+M-dwarf binaries is not always straightforward. Optical flares and soft X-ray variability are common manifestations of magnetic activity in active late-type stars \citep[e.g.][]{Gunther2020,Kowalski2024}. At the same time, Balmer emission and low-level accretion-related signatures have also been reported in detached PCEBs \citep[e.g.][]{Tappert2011,Ribeiro2013,Parsons2021}. Because these diagnostics can partly overlap, a combined analysis of high-cadence optical photometry, X-ray spectral evolution, optical spectroscopy, and broadband SED constraints is useful for assessing whether the observed variability requires an accretion-powered component or can be explained mainly by stellar magnetic activity.

The morphology of the coherent orbital waveform provides an additional diagnostic. In hot-WD PCEBs such as NN~Ser, irradiation of the companion can produce a strong reflection effect and a single-wave modulation over the orbital cycle \citep[e.g.][]{Brinkworth2006,Parsons2010}. Single-wave orbital modulations are also observed in WD pulsars, where the heating is linked to non-thermal emission from a rapidly spinning, strongly magnetized white dwarf that accelerates electrons impinging on a very close M-dwarf companion, a phenomenon first seen in AR~Sco \citep{Marsh2016,CastroSegura2025}. By contrast, a dominant signal at $2f_{\rm orb}$, together with a stable double-wave folded light curve, is the expected signature of ellipsoidal modulation from a tidally distorted companion. This distinction is especially relevant for RX~J1553.0+4457 because the WD is cool and the companion is close to filling its Roche lobe.

In this work we combine high-cadence multi-band photometry from the Burst Observer and Optical Transient Exploring System (BOOTES), public multi-sector \textit{TESS} light curves, contemporaneous \textit{Einstein Probe} (\textit{EP})/Follow-up X-ray Telescope (FXT) X-ray data, near-epoch Calar Alto Faint Object Spectrograph (CAFOS) optical spectroscopy, and archival broadband photometry. The BOOTES observations provide a multi-colour measurement of two short optical flares, including their chromatic decay and approximate band-limited energetics. The six \textit{TESS} sectors allow us to determine whether the dominant broad-band optical waveform is a stable ellipsoidal modulation at $2f_{\rm orb}$ and to build a conservative flare census in a homogeneous bandpass. The time-resolved \textit{EP}/FXT spectra trace the fading X-ray plasma after the Wide-field X-ray Telescope (WXT) detection, while the ultraviolet-to-mid-infrared SED tests whether the observed flux distribution requires a luminous disc-like, hot, or infrared-excess component. These measurements are not intended to re-derive the binary parameters; instead, they clarify the current activity state and the physical origin of the dominant optical variability.

The paper is organised as follows. In Sect.~\ref{sec:obs} we describe the observational data and reduction procedures. In Sect.~\ref{sec:results} we present the main photometric, timing, X-ray, spectroscopic, and SED results. Sect.~\ref{sec:discussion} discusses their physical implications. Finally, Sect.~\ref{sec:conclusions} summarises our conclusions. Throughout, we treat weak wind-fed accretion as compatible with a detached configuration, distinct from sustained Roche-lobe overflow, for which the present data provide no evidence.
%%%%%%%%%%%%%%%%%%%%%%%%%%%%%%%%%%%%%%%%%%%%%%%%%%%%%%%%%%%%%%%%%%%%%%%%%%%%%%%%%%%%%%%%%%%%%%%%%%%%%%%%%%%%%%%%%%%%%%%%%%%%%%%%%%%%%%%%%%%%%%%%%%%%%%%%%%%%%%%%%%%%%%%%%%%%%%%%%%%%%%%%%%%%%%%%%%%%%%%%%%%%%%%%%%%%%%%%

%--------------------------------------------------------------------
\section{Methods}
\label{sec:obs}

\subsection{BOOTES observations and flare analysis}
\label{subsec:bootes}

RX~J1553.0+4457 was monitored with the BOOTES network \citep{CastroTirado2012,CastroTirado2023NatAstron,Hu2023} of 0.6-m robotic telescopes on 2025 June 5, using the BOOTES-4/5/6/7 stations and the $g$, $r$, $i$, $z$, and clear bands. The raw detector images were processed by the standard BOOTES automatic pipeline (bias and flat correction, astrometric solution, and aperture photometry), which provides calibrated light-curve files containing Julian date, magnitude, uncertainty, and telescope/filter identifiers. We use these pipeline products directly, applying simple quality cuts to remove a few obvious outliers with large formal errors before constructing the multi-band light curves used in Sect.~\ref{subsubsec:bootes_lc}.

For the BOOTES flare analysis, the calibrated magnitudes were converted to dereddened Sloan-like AB flux densities using the same fixed foreground extinction adopted for the SED analysis, $A_V=\av$ with $R_V=3.1$. For each filter, the local quiescent level was estimated from the faintest 10\,\% of the corresponding light curve. The flare excess flux was then defined as $F_{\lambda}(t)-F_{\lambda,q}$, and the equivalent duration in a given band, ${\rm ED}_{\rm band}$, was computed using Eq.~\eqref{eq:bootes_ed},
\begin{equation}
    {\rm ED}_{\rm band} =
    \int \frac{F_{\lambda}(t)-F_{\lambda,q}}
    {F_{\lambda,q}}\,dt .
    \label{eq:bootes_ed}
\end{equation}
For the main BOOTES flare, the band-limited flare energy and peak band
luminosity were estimated using Eqs.~\eqref{eq:bootes_energy} and \eqref{eq:bootes_lpeak},
\begin{equation}
    E_{\rm band}=4\pi d^2
    \int \Delta F_{\lambda}(t)\,\Delta\lambda_{\rm band}\,dt ,
    \label{eq:bootes_energy}
\end{equation}
and
\begin{equation}
    L_{\rm peak,band}=4\pi d^2
    \Delta F_{\lambda,{\rm peak}}\,\Delta\lambda_{\rm band}.
    \label{eq:bootes_lpeak}
\end{equation}
These quantities were evaluated using the adopted distance $d=\distpc$~pc and are used only as observed-band energetics, not as bolometric flare energies.
To characterize the decay shape of the main BOOTES flare, we fitted the
post-peak excess fluxes with three empirical prescriptions, motivated by
previous empirical studies of white-light flare morphology
\citep[e.g.][]{2014ApJ...797..121H,Davenport2014}. These were the simple power-law, exponential, and generalized power-law forms given in Eqs.~\eqref{eq:pl_decay}, \eqref{eq:exp_decay}, and \eqref{eq:gpl_decay}. The simple power-law form is
\begin{equation}
    \Delta F_{\lambda}(t)
    = A\left(\frac{t-t_{\rm p}}{1\,{\rm d}}\right)^{-\alpha},
    \label{eq:pl_decay}
\end{equation}
the exponential form is
\begin{equation}
    \Delta F_{\lambda}(t)
    = A\exp\left[-\frac{t-t_{\rm p}}{\tau}\right],
    \label{eq:exp_decay}
\end{equation}
and the generalized power-law form is
\begin{equation}
    \Delta F_{\lambda}(t)
    = A\left(1+\frac{t-t_{\rm p}}{\tau}\right)^{-\alpha}.
    \label{eq:gpl_decay}
\end{equation}
Here $t_{\rm p}$ is the observed flare-peak time in each band, $A$ is a normalization, $\tau$ is a characteristic decay timescale, and $\alpha$ describes the late-time decay slope. These functions are used only as empirical descriptions of the observed post-peak flux evolution.

\subsection{\textit{TESS} light-curve extraction}

We analysed public \textit{TESS} full-frame-image cutouts from Sectors~23, 24, 50, 51, 77, and 78, obtained via \texttt{TESScut} \citep{Ricker2015,Brasseur2019}. The cutouts were reduced in a uniform way using \texttt{lightkurve} \citep{Lightkurve2018}, while all time-series analysis was carried out with routines from \texttt{astropy} \citep{Astropy2022}. Because our data products are local cutouts rather than standard mission-pipeline target light curves, we re-extracted the photometry consistently for all sectors.

For each cutout, we defined a source aperture using \texttt{lightkurve}'s threshold mask with a $3\sigma$ threshold and the cutout center as the reference pixel. If the threshold mask failed or returned an empty aperture, we adopted a fallback $3\times3$ central mask. Aperture photometry was then extracted from the target-pixel data, and cadences with non-zero quality flags were rejected whenever such flags were available. We further removed non-finite measurements and discarded sectors with too few valid cadences.

Each sector light curve was normalized by its median flux, followed by a symmetric $5\sigma$ clipping to suppress isolated outliers. Long-timescale variations were then removed sector by sector using \texttt{lightkurve.flatten} with a window length of 401 cadences. After detrending, we applied a second $5\sigma$ clipping to the flattened light curve and converted the final flux series to relative units, $f_{\rm rel}=f/\tilde{f}-1$, where $\tilde{f}$ denotes the sector median before detrending. The cleaned sector light curves were sorted in time and used for the combined period search as well as for the sector-by-sector folded-light-curve comparison.

\subsection{\textit{TESS} flare search and completeness assessment}
\label{subsec:tess_flare_search}

To construct a homogeneous \textit{TESS} flare sample, we started from the cleaned sector light curves described above and first removed the dominant orbital waveform. For each sector, we fitted a fixed-period two-harmonic model at the literature orbital period and searched the residual relative-flux series for positive excursions above the local scatter. Candidate events were identified from contiguous groups of points exceeding a seed threshold and were then re-measured after conservative expansion of their temporal boundaries. This conservative identification of white-light flares from detrended residual light curves follows the general approach widely used in space-based flare studies \citep[e.g.,][]{Davenport2014}.

For each candidate we measured the peak relative amplitude, event duration, and equivalent duration (ED) using Eq.~\eqref{eq:tess_ed} \citep{Gershberg1972},
\begin{equation}
{\rm ED} = \int \left(\frac{f}{f_{\rm q}} - 1\right)\,dt,
    \label{eq:tess_ed}
\end{equation}
where $f_{\rm q}$ is the local quiescent level after subtraction of the orbital modulation. We then defined the final flare sample by applying conservative cuts in peak signal-to-noise ratio, peak relative amplitude, duration, and ED, and by merging nearby candidates separated by short temporal gaps before re-measuring the merged event.

The selection function was quantified with injection--recovery tests performed directly on the residual \textit{TESS} light curves, following the general logic commonly adopted in modern flare-catalog and completeness analyses \citep[e.g.,][]{Gunther2020}. Synthetic flares were injected over a grid of input ED values while avoiding time windows containing real flare candidates. For each trial we recorded both the raw detection outcome and the final clean-sample recovery outcome. Because the scientific sample used in this work is the final cleaned sample, the adopted completeness thresholds were defined from the clean-recovery curve rather than from the raw-detection curve.

At high ED, where the measured clean-recovery fraction is not guaranteed to remain strictly monotonic for a finite set of trials, we used the monotonic envelope of the clean-recovery curve to derive the characteristic completeness thresholds. In this notation, ${\rm ED}_{x}$ denotes the equivalent-duration threshold at which $x$ per cent of the injected synthetic flares are recovered in the final clean sample. It is based on the \textit{TESS} ED definition in Eq.~\eqref{eq:tess_ed}, which is conceptually analogous to the band-specific BOOTES ${\rm ED}_{\rm band}$ in Eq.~\eqref{eq:bootes_ed}; however, ${\rm ED}_{80}$ is a completeness threshold, not the equivalent duration of an individual flare in a particular filter. In the analysis below we adopt ${\rm ED}_{80}$ as the main working threshold for the cumulative flare-frequency analysis. Additional completeness and robustness checks are presented in Appendix~\ref{app:completeness}.
%%%%%%%%%%%%%%%%%%%%%%%%%%%%%%%%%%%%%%%%%%%%%%%%%%%%
\subsection{\textit{EP} data analysis}

\textit{EP}/WXT detected an X-ray transient at the time of 2025-06-05 18:51:47 (UTC) (trigger ID 01709177873, CMOS 42), which is reported as a likely stellar flare associated with RX~J1553.0+4457 \citep{Lian2025GCN40635}. The WXT source is positioned at right ascension $238.261834^\circ$ and declination $+44.968370^\circ$, with a detection significance of $6.95\sigma$. \textit{EP}/WXT continued to observe the source with its CMOS 14 and CMOS 37 in the $0.5$--$4.0$\,keV band. $198.6\ \mathrm{s}$ after the WXT trigger, \textit{EP}/FXT initiated follow-up observations with its two modules, FXT-A and FXT-B, in the $0.3$--$10.0$\,keV band, pointing the source at right ascension $238.238993^\circ$ and declination $+44.957061^\circ$, consistent with the location of the WXT transient. Because the FXT observation began $\sim 200\ \mathrm{s}$ after the WXT trigger, the FXT data cover only the declining phase of the flare; they do not constrain the onset time and likely miss the flare peak. The FXT observation spans a total duration of $8530.3\ \mathrm{s}$, from 2025-06-05 18:55:05.6 to 21:17:15.9 (UTC), with an exposure of $5746.4\ \mathrm{s}$ and a gap of $2783.9\ \mathrm{s}$ caused by occultation of the target during the intervening satellite orbit.

The dataset consists of two cleaned FXT event files from the two modules together with the corresponding source and background spectral products and the standard response files, produced within the standard FXT data-processing framework \citep{Zhao2025FXTData}. In the present work, the \textit{EP} data are used primarily for a time-resolved spectral characterization of the rapid X-ray decline recorded during the FXT coverage.

We use the \textsc{Xselect} tool to extract the pileup-corrected \textit{EP}/FXT spectrum of the full observation as a time-averaged reference spectrum to guide the time-resolved fitting. We then extracted a sequence of time-resolved spectra covering the observed decline phase, binned to achieve a minimum photon count of $2000$ in each slice. Event times were converted to absolute UTC values using the event-file timing keywords, so that each spectral bin could be placed on a common time axis and compared directly with the \textit{EP} trigger time in Fig.~\ref{fig:ep_fxt_context}. 

To illustrate the spectral features of thermal plasma during this stellar flare,
all FXT spectra were fitted in \textsc{XSPEC} \citep{Arnaud1996} using the same
three-temperature Astrophysical Plasma Emission Code (\textsc{APEC}) optically
thin thermal-plasma model \citep{Smith2001},
\texttt{tbabs*(apec+apec+apec)}, with the first component \texttt{tbabs}
\citep{tbabs} accounting for Galactic absorption. For the absorbing column, we
adopted $N_{\rm H} = 8.83 \times 10^{18}\,{\rm cm^{-2}}$, obtained from the
full-observation reference fit in XSPEC. At such a small column, absorption has
negligible leverage on the $0.3$--$10.0$\,keV FXT spectra, so $N_{\rm H}$ was
fixed rather than fitted for the time-resolved spectra. We also fixed the
abundance to $1.0$ times the solar value \citep{ANDERS1989197} and the redshift
to $0$, as appropriate for a Galactic source, allowing only the temperatures
and normalizations of the three components to vary freely. The normalizations
were used as proxies for the corresponding emission measures. We performed the
fits using the C-statistic with background in XSPEC \citep{Cash1979}, fitting
FXT-A and FXT-B spectra simultaneously with identical model parameters. For
each time-resolved spectrum we recorded the best-fitting temperatures and
emission measures of the three components, and used the model \texttt{cflux} to
derive the unabsorbed $0.3$--$10.0$\,keV flux $F_X$.

%----------------

\subsection{\textit{TESS} period analysis}

We searched for periodic variability with the Lomb--Scargle (LS) formalism \citep{Lomb1976,Scargle1982,VanderPlas2018} as implemented in \texttt{astropy.timeseries} \citep{Astropy2022}. The cleaned light curves from all available \textit{TESS} sectors were concatenated, sorted in time, and analysed on a frequency grid spanning $0.01~\mathrm{d^{-1}}$ to $0.9$ times the Nyquist frequency defined from the median cadence, with an oversampling factor of 10.

Because the orbital period of RX~J1553.0+4457 is already known from phase-resolved spectroscopy, our goal here is not an independent blind period determination, but to test whether the dominant \textit{TESS} modulation is consistent with the expected first harmonic of the orbital signal. We therefore examined both the full LS periodogram and a local frequency window centred on $2f_{\rm orb}$, adopting a half-width of $0.5~\mathrm{d^{-1}}$.

To assess the local significance against correlated variability, we first fitted and subtracted a fixed two-harmonic model at the literature orbital frequency. The residual broadband variability was then described with the power-law plus white-noise floor in Eq.~\eqref{eq:rednoise_model},
\begin{equation}
P(f)=Af^{-\alpha}+C.
    \label{eq:rednoise_model}
\end{equation}
From this model we generated 1000 Timmer--Koenig red-noise surrogate light curves sampled at the real timestamps \citep{Timmer1995,Vaughan2005}, and for each realization we recorded the maximum LS power within the local search band around $2f_{\rm orb}$. The empirical distribution of these maxima defines the local confidence levels used for the significance assessment of the harmonic peak. A moving-block bootstrap was also performed as a secondary robustness check, but the red-noise surrogate test is adopted as the primary significance assessment.

%----------------
\subsection{Optical spectroscopy}
\label{subsec:optical_spectroscopy_methods}

Optical spectroscopy of RX~J1553.0+4457 was obtained on 2025 June 9 with the Calar Alto Faint Object Spectrograph (CAFOS; \citealt{CortesContreras2023}) mounted on the 2.2\,m Calar Alto telescope. The target was observed with the B100, G100, and R100 grisms, with two 300\,s exposures obtained in each grism setting. The observing sequence also included HgHeRb arc lamps, bias frames, dome flats, and observations of the spectrophotometric standard star BD\,+33\,2642 for flux calibration.

\begin{figure*}[!ht]
    \centering
    \includegraphics[width=0.90\textwidth]{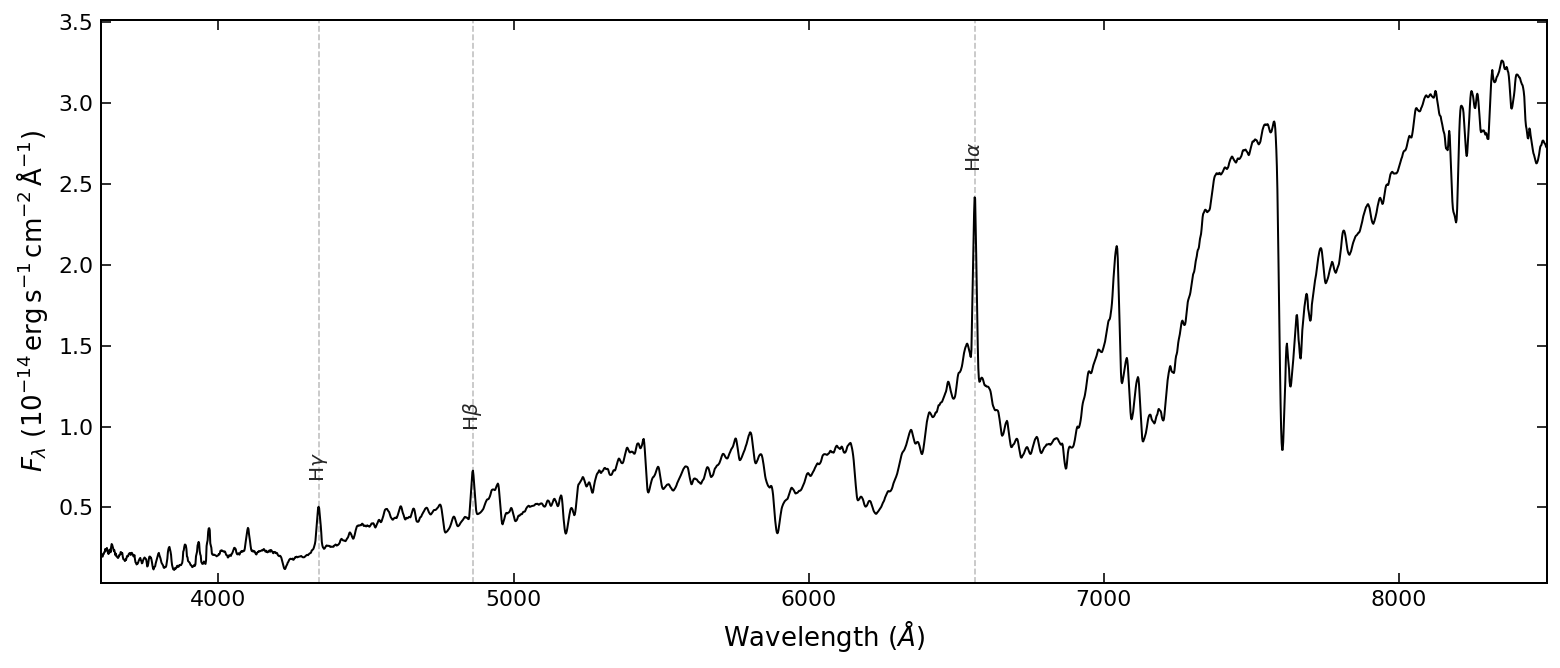}
    \caption{
    Displayed CAFOS optical spectrum of RX~J1553.0+4457 over the full merged B100, G100, and R100 wavelength range, taken on 2025 June 9. Dashed vertical lines mark the laboratory wavelengths of the main Balmer lines discussed in the text. The plotted spectrum is shown in units of $10^{-14}\,\mathrm{erg\,s^{-1}\,cm^{-2}}\,\text{\AA}^{-1}$ and has been mildly cleaned and smoothed for display only; the measurements in Table~\ref{tab:cafos_line_measurements} are performed on the unsmoothed merged spectrum.
    }
    \label{fig:spec}
\end{figure*}

The two extracted one-dimensional spectra in each grism setting were interpolated onto a common linear wavelength grid and combined using inverse-variance weighting. The B100, G100, and R100 segments were then merged by rescaling the moving segment to the reference segment using the median flux ratio measured in the overlap region; overlap pixels were again combined using inverse-variance weighting. A mildly cleaned and smoothed version of the merged spectrum is used only for display in Fig.~\ref{fig:spec}.

Emission-line measurements were performed on the unsmoothed merged spectrum. For each line, we fitted a first-order local continuum using two adjacent continuum windows and integrated the continuum-subtracted profile over a fixed line window. The emission equivalent width was defined as positive for emission using Eq.~\eqref{eq:ew_emission},
\begin{equation}
    W_{\rm em} =
    \int \left(\frac{F_\lambda}{F_{\rm cont}} - 1\right)\,d\lambda .
    \label{eq:ew_emission}
\end{equation}
Uncertainties were estimated by Monte Carlo resampling of the flux array using the propagated spectral uncertainties, with the local continuum refitted in each realization. The resulting values are intended as single-epoch, low-to-moderate-resolution descriptive measurements. Because the three grism settings were obtained sequentially and then rescaled in their overlap regions, we do not use the spectrum for phase-resolved line-profile diagnostics or for physical interpretation of Balmer line ratios.

\subsection{Spectral energy distribution modelling}

Given the independently established WD+M-dwarf nature of RX~J1553.0+4457 \citep{Liu2025}, we used the SED primarily to test whether the available ultraviolet-to-mid-infrared photometry is consistent with the two stellar photospheres and whether any additional hot, disc-like, or infrared-excess component is required. The fit was therefore restricted to a WD+M-dwarf template combination, rather than being used as a fully independent determination of the stellar atmospheric parameters.

Photometry was assembled and inspected with the Virtual Observatory SED Analyzer (\textsc{VOSA}; \citealt{Bayo2008}). The archival photometric constraints were drawn from the \textit{Galaxy Evolution Explorer} (\textit{GALEX}; \citealt{Martin2005,morrissey2007}), the Sloan Digital Sky Survey (SDSS; \citealt{Alam2015}), the AAVSO Photometric All-Sky Survey (APASS; \citealt{Henden2016}), \textit{Gaia} Data Release 3 (DR3; \citealt{GaiaCollaboration2023}), the Two Micron All Sky Survey (2MASS; \citealt{Skrutskie2006}), and the \textit{Wide-field Infrared Survey Explorer} (\textit{WISE}) and \textit{Near-Earth Object Wide-field Infrared Survey Explorer} (\textit{NEOWISE}) missions \citep{wright2010,Mainzer2011}. Known problematic and synthetic photometric entries were excluded, and the final fitted dataset comprises 19 photometric constraints: 18 detections and one $3\sigma$ upper limit, corresponding to the \textit{GALEX} far-ultraviolet (FUV) point.

The distance was not fitted in the SED analysis. We adopted the \textit{Gaia} Early Data Release 3/Bailer-Jones distance \citep{GaiaCollaboration2021,BailerJones2021}, $d=\distpc~\mathrm{pc}$, and entered this value as a fixed distance in \textsc{VOSA}. We also adopted a low fixed foreground extinction of $A_V=\av$ with $R_V=3.1$ \citep{Cardelli1989}. This value is consistent with the Bayestar19 three-dimensional dust map along this high-latitude line of sight, which gives $E(B-V)=0.0088$ and therefore $A_V=0.027$ for $R_V=3.1$ \citep{Green2019}. The extinction was used as a fixed input for dereddening the photometry in \textsc{VOSA}, rather than being treated as a fitted parameter.

We explored binary combinations of Koester WD atmospheres \citep{Koester2010} and BT--Settl (CIFIST2011) companion models \citep{Allard2012}. Because broadband photometry alone cannot reliably determine the WD surface gravity, we adopted a physically constrained reference solution with the WD fixed at $\log g=\wdlogg$. This solution has a Koester WD with $T_{\rm eff}=\wdteff$~K and a BT--Settl companion with $T_{\rm eff}=\mdteff$~K and $\log g=\mdlogg$. Additional trial fits in which the WD surface gravity was allowed to move to higher grid values were used only as robustness checks. These trials gave slightly lower formal $\chi^2$ values but drove the WD solution towards the upper edge of the allowed $\log g$ range, while leaving the companion temperature essentially unchanged.

After the \textsc{VOSA} fit, the fixed two-component templates were used in a post-\textsc{VOSA} scaling check to inspect residuals and to generate the SED figure. No additional extinction parameter was introduced in this step, and the sampled scale and jitter nuisance parameters were not used as independent physical constraints on the stellar components.

%%%%%%%%%%%%%%%%%%%%%%%%%%%%%%%%%%%%%%%%%%%%%%%%%%%%%%%%%%%%%%%%%%%%%%%%%%%%%%%%%%%%%%%%%%%%%%%%%%%%%%%%%%%%%%%%%%%%%%%%%%%%%

%-----------------------------------------------------------------

\section{Results}
\label{sec:results}

\subsection{BOOTES light curves}
\label{subsubsec:bootes_lc}

\begin{figure*}[htbp]
    \centering
    \includegraphics[width=0.90\textwidth]{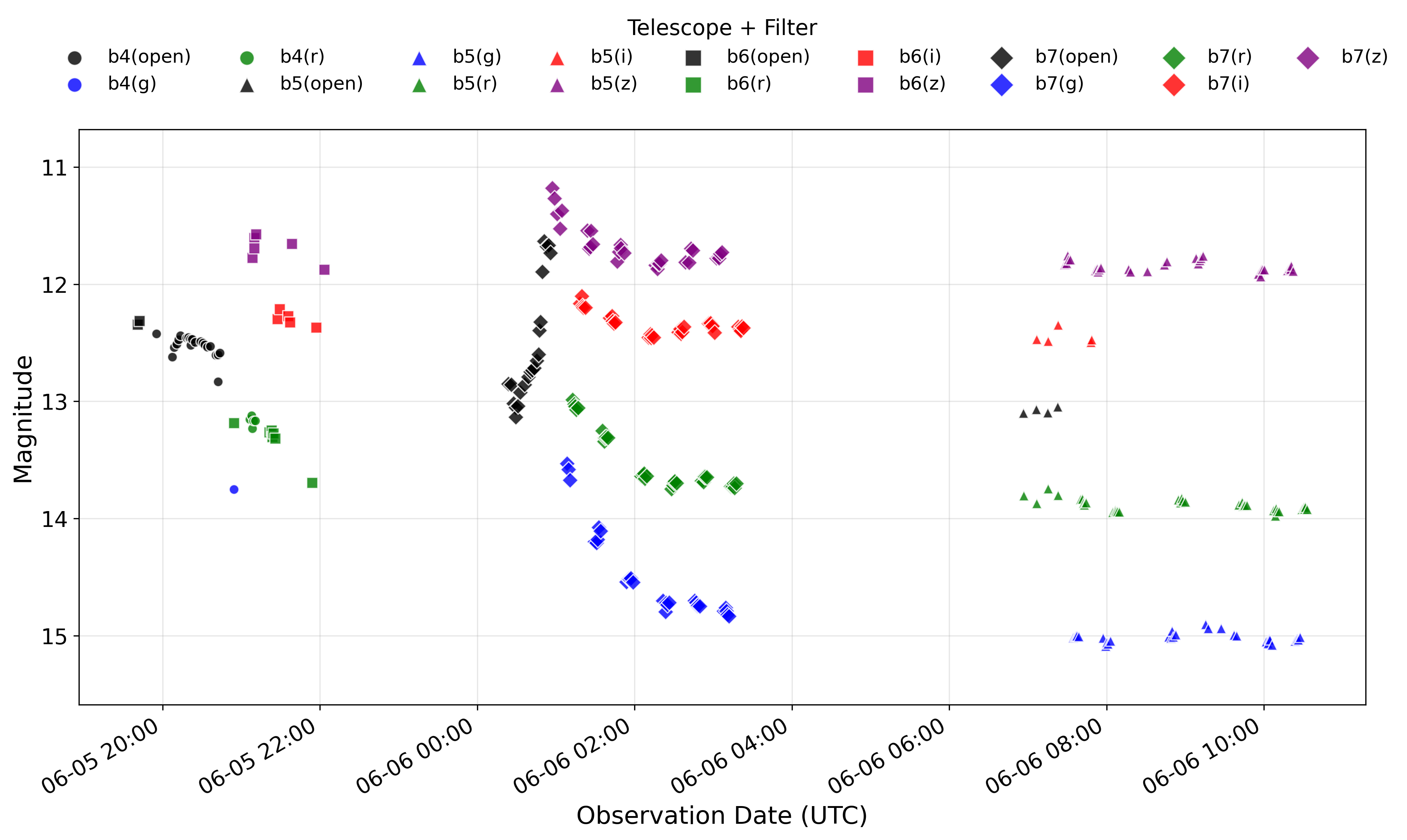}
    \caption{
    Multi-band BOOTES light curve of RX~J1553.0+4457 from 2025-06-05 19:40:51 to 2025-06-06 10:32:56~UTC (total span $\simeq0.62$~d).
    The plot shows 288 individual measurements obtained with four ($b4$--$b7$) of the seven BOOTES telescopes worldwide in the $g$, $r$, $i$, $z$, and open (unfiltered) bands.
    Different colours and symbols indicate distinct telescope--filter combinations as indicated in the legend.
    Two bright flares are clearly seen, separated by $\sim3$~h, superimposed on a slowly varying baseline.
    }
    \label{fig:bootes_lightcurve}
\end{figure*}

Figure~\ref{fig:bootes_lightcurve} shows the BOOTES multi-band monitoring obtained on 2025 June 5/6. The light curve contains two prominent short-lived flares separated by $\sim3$~h, with amplitudes of $\sim1$--1.5~mag above the local quiescent level and decay timescales of tens of minutes.

\begin{figure}[!ht]
    \centering
    \includegraphics[width=0.48\textwidth]{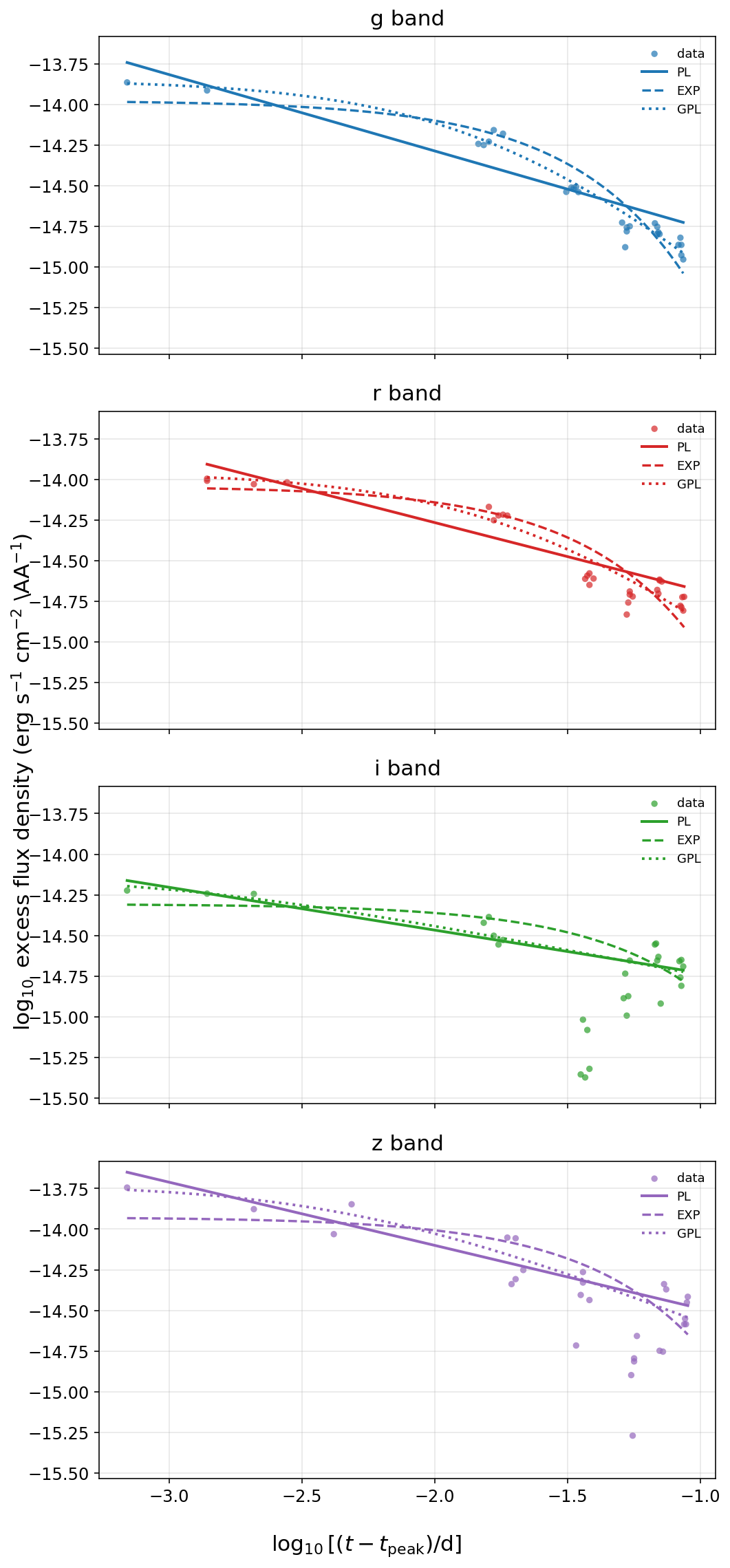}
    \caption{
    Comparison of analytic prescriptions for the decay of the main BOOTES flare.
    Each panel shows $\log_{10}[(t-t_{\rm peak})/{\rm d}]$ versus the
    $\log_{10}$ excess flux density in one of the $g$, $r$, $i$, and $z$ bands,
    where $t_{\rm peak}$ is the observed flare maximum in the corresponding band.
    The points indicate the post-peak data used in the fits, while the solid,
    dashed, and dotted curves show the best-fitting simple power-law,
    exponential, and generalized power-law models, respectively.
    All fits are performed in log-flux space using the same baseline convention
    and post-peak selection procedure described in Sect.~\ref{subsec:bootes}.
    Because the time axis is referenced to the band-specific flare maximum, this
    figure is intended to compare decay-model shapes rather than absolute
    inter-band timing offsets.
    }
    \label{fig:bootes_modelcomp}
\end{figure}

The resulting flare-decay fits and model comparisons for the main flare are shown in Fig.~\ref{fig:bootes_modelcomp}. Using the generalized power-law (GPL) parameterization defined in Sect.~\ref{subsec:bootes}, we obtain
$\alpha_g = 1.41^{+0.09}_{-0.11}$,
$\alpha_r = 1.24^{+0.06}_{-0.10}$,
$\alpha_i = 0.31^{+0.04}_{-0.05}$, and
$\alpha_z = 0.65^{+0.04}_{-0.08}$,
where the quoted uncertainties conservatively enclose both the Monte Carlo photometric perturbations and the parameter-selection robustness grid. The decay is therefore clearly chromatic: the $g$- and $r$-band  indices are substantially larger than those in the red bands,  indicating faster fading at shorter wavelengths. The ordering is  not strictly monotonic with wavelength, however, as  $\alpha_z > \alpha_i$ despite the $z$ band being the redder of  the two. The very shallow $i$-band value, $\alpha_i=0.31^{+0.04}_{-0.05}$, should not be interpreted as a separate dynamical timescale. In this band the flare contrast is lower and the quiescent contribution from the late-type donor is more important than in the bluer filters. The $i$ band may also be more sensitive to bandpass-dependent effects, including partial filling of broad molecular absorption features such as TiO as the flare continuum cools. We therefore regard the low $\alpha_i$ mainly as evidence for chromatic flare evolution and red-band dilution, rather than as a direct bolometric decay index.

This wavelength dependence is qualitatively consistent with a cooling flare spectrum, in which the blue bands are more strongly weighted toward higher-temperature emission that decays more rapidly after the peak \citep{Doyle2019,Kowalski2024}. Using the observed-band energy definitions given in Sect.~\ref{subsec:bootes}, the main-flare equivalent duration ranges from $\sim3.4$--$3.7$~ks in $i$ and $z$ to $\sim48.9$~ks in $g$, corresponding to band-limited energies of $(1.33$--$5.08)\times10^{34}$~erg and peak band luminosities of $(2.01$--$3.33)\times10^{30}$~erg\,s$^{-1}$ for the adopted distance (Table~\ref{tab:flare_summary}).

For model comparison, we fitted simple power-law (PL), exponential (EXP), and GPL prescriptions to the same post-peak segments. Exponential decays are disfavoured in all bands. GPL provides the best description in the blue bands ($g$ and $r$), while in the redder bands the improvement becomes weaker: in $z$ it is only marginal, and in $i$ a simple PL already provides an adequate description.

%------
\begin{table*}[htbp]
\centering
\caption{Main-flare decay and approximate band-limited energetics from the BOOTES data. The equivalent durations, energies, and peak luminosities are measured in the observed $g$, $r$, $i$, and $z$ bands after converting the calibrated magnitudes to dereddened Sloan-like AB flux densities. The quoted energies are band-limited values, not bolometric flare energies. The quoted uncertainties combine Monte Carlo photometric perturbations with the robustness-grid variations in the baseline and fitting-window choices.}
\label{tab:flare_summary}

\setlength{\tabcolsep}{4pt}
\renewcommand{\arraystretch}{1.4}
\begin{tabular}{lcccc}
\toprule
Filter & $\alpha_{\rm GPL}$ & ED [ks] & $E_{\rm band}$ [$10^{34}$ erg] & $L_{\rm peak,band}$ [$10^{30}$ erg\,s$^{-1}$] \\
\midrule
$g$ & $1.41^{+0.10}_{-0.11}$ & $48.9^{+1.7}_{-1.7}$ & $5.08^{+0.17}_{-0.13}$ & $3.24^{+0.17}_{-0.12}$ \\
$r$ & $1.24^{+0.10}_{-0.10}$ & $18.9^{+1.1}_{-0.5}$ & $3.23^{+0.16}_{-0.06}$ & $2.40^{+0.07}_{-0.04}$ \\
$i$ & $0.31^{+0.06}_{-0.06}$ & $3.4^{+0.5}_{-0.3}$ & $1.65^{+0.20}_{-0.11}$ & $2.01^{+0.21}_{-0.12}$ \\
$z$ & $0.65^{+0.09}_{-0.09}$ & $3.7^{+0.4}_{-0.2}$ & $1.33^{+0.10}_{-0.06}$ & $3.33^{+0.11}_{-0.05}$ \\
\bottomrule
\end{tabular}
\end{table*}

%------

The occurrence of two BOOTES flares during a single night of targeted follow-up should not be taken as representative of the long-term flare rate.  As a historical context, we queried the public Zwicky Transient Facility (ZTF) archive through the NASA/IPAC Infrared Science Archive \citep{Teplitz2018,Bellm2019,Masci2019} in the available $g$, $r$, and $i$ filters and applied conservative quality cuts.  The resulting light curve contains 2903 usable measurements over a 7.56 yr baseline.  Using band-dependent quiescent baselines, we find four ZTF-sampled episodes brighter than the baseline by $\Delta m\ge0.50$ mag, corresponding to a sampled-event rate of $0.53~{\rm yr^{-1}}$. Treating these sparse detections as Poisson counts \citep{Gehrels1986} gives a one-sided 95 per cent upper limit of $\lesssim1.2~{\rm yr^{-1}}$ for such ZTF-sampled large-amplitude brightenings.  The $i$-band data were also checked, but they do not provide an independent large-brightening constraint because the maximum measured $i$-band brightening is only 0.21 mag.  Because the ZTF cadence is sparse, this estimate is not a direct flare rate; it only shows that comparable large optical brightenings are not continuously present in the long-term record.  We therefore use the BOOTES events as evidence for episodic activity during the follow-up night, rather than as a measurement of the typical duty cycle.

%-----%-----%-----%-----%-----%-----%-----%-----%-----%-----

%-----%-----%-----%-----%-----%-----%-----%-----%-----%-----%-----%-----
\subsection{\textit{TESS} orbital modulation}
\label{subsec:period_analysis}

\begin{figure}[htbp]
\centering
\includegraphics[width=0.49\textwidth]{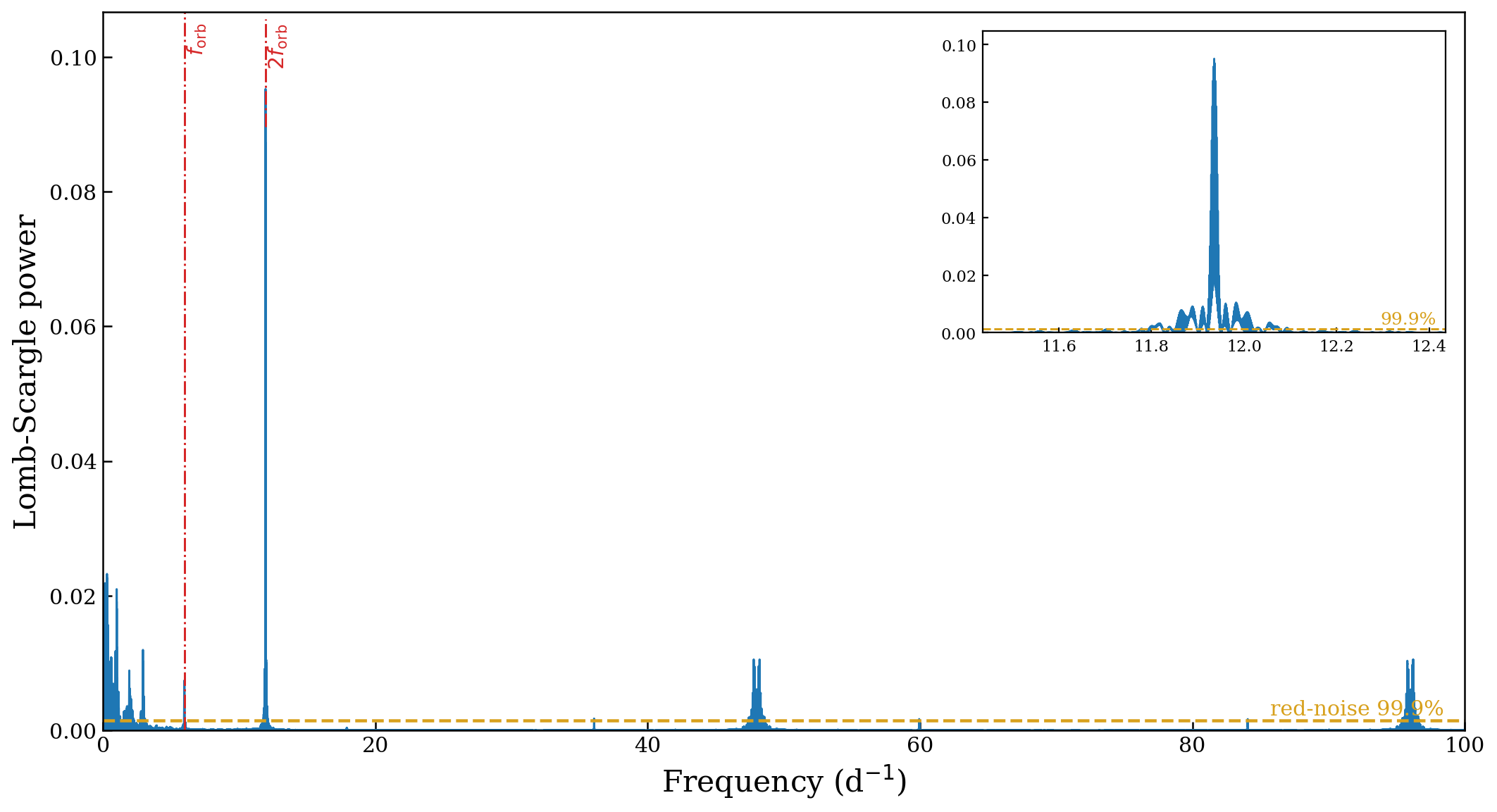}

\vspace{0.35cm}

\includegraphics[width=0.49\textwidth]{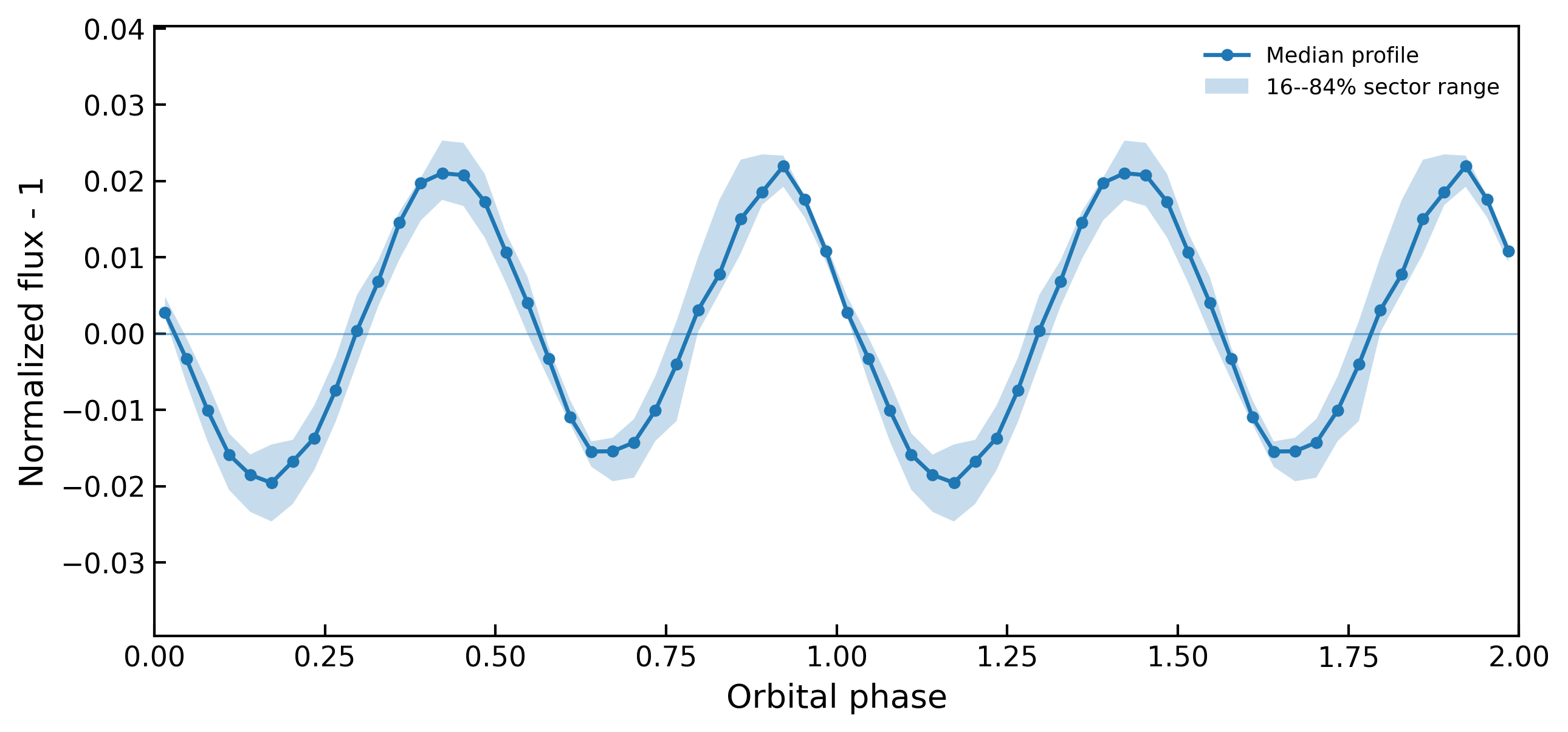}
\caption{
Timing diagnostics from the multi-sector \textit{TESS} light curve of RX~J1553.0+4457.
Upper panel: Lomb--Scargle periodogram of the combined light curve. The red dash-dotted lines mark $f_{\rm orb}$ and $2f_{\rm orb}$, and the inset zooms in on the local frequency window around $2f_{\rm orb}$. The dashed horizontal line indicates the 99.9 per cent local red-noise threshold used for the significance test around the first harmonic.
Lower panel: sector-median waveform after folding the six \textit{TESS} sectors on the spectroscopic orbital period.
The solid curve gives the median profile, and the shaded band gives the 16th--84th percentile range across the sector-level profiles.
The coincidence of the dominant periodogram peak with $2f_{\rm orb}$, together with the stable double-wave folded morphology, supports an ellipsoidal interpretation of the broad-band orbital modulation.
}
\label{fig:ls_periodogram}
\end{figure}

The combined multi-sector \textit{TESS} light curve shows a dominant
Lomb--Scargle (LS) peak at $f_{\rm peak}=11.935709~{\rm d}^{-1}$
(Fig.~\ref{fig:ls_periodogram}), corresponding to $P_{\rm peak}=0.083782~{\rm d}$. This frequency is fully consistent with twice the spectroscopic orbital frequency reported by \citet{Liu2025}. For $P_{\rm orb}=0.16756456~{\rm d}$, the expected first harmonic is $2f_{\rm orb}=11.935698~{\rm d}^{-1}$, giving $\Delta f = +1.1 \times 10^{-5}~{\rm d}^{-1}$. This offset is far below the formal frequency resolution of the combined dataset. We therefore identify the \textit{TESS} peak with the first harmonic of the known orbital modulation.

The physical significance of the $2f_{\rm orb}$ peak is that it points to ellipsoidal modulation as the dominant broad-band optical waveform. A tidally distorted companion produces two maxima and two minima per orbital cycle, so the strongest Fourier power is expected at the first harmonic rather than at $f_{\rm orb}$ itself \citep[e.g.][]{Morris1985,MorrisNaftilan1993}. The \textit{TESS} periodogram therefore does more than recover the spectroscopic orbital solution: it shows that the stable optical modulation is dominated by a double-wave geometric component rather than by a single-wave irradiation or hot-face component.

The local red-noise significance test gives the same result. In the $\pm0.5\,{\rm d^{-1}}$ window around $2f_{\rm orb}$, the observed peak has a Lomb--Scargle power of 0.09516, well above the 99.9 per cent local red-noise threshold of 0.00141. The corresponding empirical local false alarm probability is $9.99\times10^{-4}$ for the red-noise surrogate test and $2.00\times10^{-3}$ for the moving-block bootstrap cross-check. Over the low-frequency range most relevant for the orbital signal, the periodogram is dominated by narrow coherent structure rather than by a broad accretion-like red-noise continuum.

To test whether this harmonic signal is also stable in phase space, we constructed sector-level folded profiles. Each \textit{TESS} sector was folded separately on the spectroscopic orbital period and binned in orbital phase. At each phase bin we then took the median of the six sector profiles. The shaded interval in Fig.~\ref{fig:ls_periodogram} is the corresponding 16th--84th percentile range across the six sectors. We use this interval only as a robust description of the central sector-to-sector scatter, not as a formal uncertainty on the mean profile.

This phase-domain check supports the same interpretation as the frequency-domain analysis. The sector-median profile shows a repeatable double-wave morphology when folded on $P_{\rm orb}$, while the sector-to-sector scatter remains modest compared with the full modulation amplitude. Together with the LS peak at $2f_{\rm orb}$, this indicates that the dominant \textit{TESS} modulation is phase-locked to the ellipsoidal first harmonic of the orbital cycle.

Narrow isolated peaks are also present at much higher frequencies near $132.07$ and $155.94~{\rm d}^{-1}$. These are consistent with sampling aliases of the dominant modulation for the 10-min cadence, that is, $f_{\rm alias} \simeq f_{\rm s} \pm f_{\rm peak}$ with $f_{\rm s}=144~{\rm d}^{-1}$, and are therefore not interpreted as additional physical periodicities.

%-----

\subsection{\textit{TESS} flare census}
\label{subsec:tess_flare_census}

The automatic search of the six public \textit{TESS} sectors yielded a conservative clean sample of 13 flares over a total effective observing time of 123.98~d. These events were measured from the orbital-model-subtracted residual light curves using the selection and merging procedure described in Sect.~\ref{subsec:tess_flare_search}. The individual clean flares, equivalent durations, and approximate energies are listed in Table~\ref{tab:tess_flare_catalogue}; the final selection thresholds are summarized in the table footnote.

From the clean-sample injection--recovery analysis, we obtained completeness thresholds of ${\rm ED}_{50}=21.7$\,s and ${\rm ED}_{80}=40.4$\,s, while ${\rm ED}_{90}$ was not reached within the explored ED range. We therefore adopt ${\rm ED}_{80}$ as the main threshold for subsequent flare-frequency analysis. The corresponding completeness curve should be interpreted as the selection function of the final clean flare sample, not as a pure raw-detection efficiency. The detailed recovery tests and robustness checks are given in Appendix~\ref{app:completeness}.

For broader context, we converted the RX~J1553.0+4457 \textit{TESS} flare sample to an approximate bolometric energy scale and compared it with the M-dwarf flare catalogue of \citet{Gunther2020}, which provides a useful population baseline for active low-mass stars observed with \textit{TESS}. For each RX~J1553.0+4457 event, the flare equivalent duration was measured from the orbital-model-subtracted relative-excess light curve. The corresponding \textit{TESS}-band flare energy was then estimated from the standard equivalent-duration relation in Eq.~\eqref{eq:tess_energy}
\citep{Gershberg1972,Davenport2014},
\begin{equation}
    E_{\rm TESS} = {\rm ED}\,L_{{\rm q},T},
    \label{eq:tess_energy}
\end{equation}
where $L_{{\rm q},T}$ is the quiescent luminosity of the source in the
\textit{TESS} band. In the updated calibration, we used $L_{{\rm q},T}=1.65\times10^{31}\,{\rm erg\,s^{-1}}$, obtained by integrating the scaled BT--Settl companion model adopted in the SED interpretation ($T_{\rm eff}=3200$\,K, $\log g=4.5$) over the \textit{TESS} response at the fixed distance $d=36.833$\,pc. The resulting \textit{TESS}-band flare energies span $E_{\rm TESS}=5.0\times10^{32}$--$4.2\times10^{33}$~erg for the 13 clean events. To place these energies on an approximate bolometric scale for comparison, we adopted $E_{\rm bol,app}=4.8\,E_{\rm TESS}$, following the convention used for Proxima Cen \textit{TESS} flares by \citet{Vida2019}. This gives $E_{\rm bol,app}=2.4\times10^{33}$--$2.0\times10^{34}$~erg. The factor of 4.8 should be regarded as an approximate bandpass correction rather than a source-specific calorimetric measurement; more detailed treatments derive bolometric flare energies by convolving an assumed flare spectrum with the instrumental response function \citep{Shibayama2013,Gunther2020}. These bolometric energies are therefore used only for population-level comparison and should be interpreted as order-of-magnitude estimates rather than direct multi-band calorimetric measurements.

\begin{figure}[t]
\centering
\includegraphics[width=\hsize]{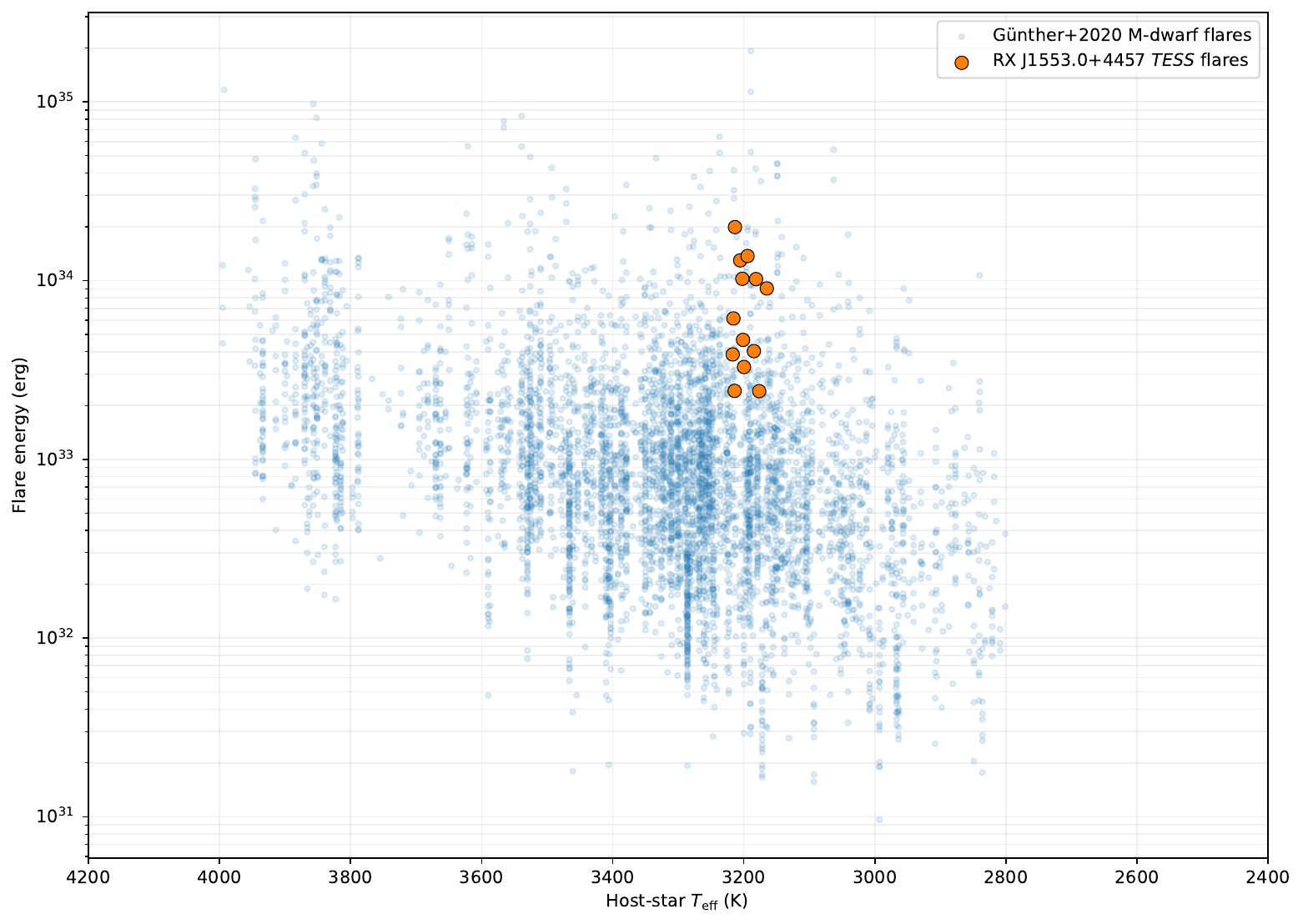}
\caption{
Approximate bolometric flare energies of the RX~J1553.0+4457
\textit{TESS} events compared with literature M-dwarf flares from
\citet{Gunther2020}, shown as flare energy versus host-star effective
temperature. The RX~J1553.0+4457 points use the updated \textit{TESS}-band
energy calibration and are converted for comparison using
$E_{\rm bol,app}=4.8\,E_{\rm TESS}$. The BOOTES flares are not shown in
this comparison because their multi-band optical energies are not derived in
the same homogeneous \textit{TESS}-band framework.
}
\label{fig:tess_flare_context}
\end{figure}

Figure~\ref{fig:tess_flare_context} shows that the 13 \textit{TESS} flares identified here fall within the energetic range occupied by active M-dwarf flares, although they lie toward the upper part of the comparison distribution at $T_{\rm eff}\simeq3200$~K. This agrees with the broader \textit{TESS} view that active late-type stars can produce energetic optical flares across a wide range of spot and rotational properties \citep{Doyle2019,Kowalski2024}. We use this comparison as a population-level reference for the \textit{TESS} flare sample. The BOOTES events are kept separate because their multi-band optical energies are not derived in the same homogeneous \textit{TESS}-band framework.

%-----%-----%-----%-----%-----%-----%-----%-----%-----%-----%-----%-----%-----%-----%-----

\subsection{\textit{EP} data results}

\begin{figure}[ht]
    \centering
    \includegraphics[width=\columnwidth]{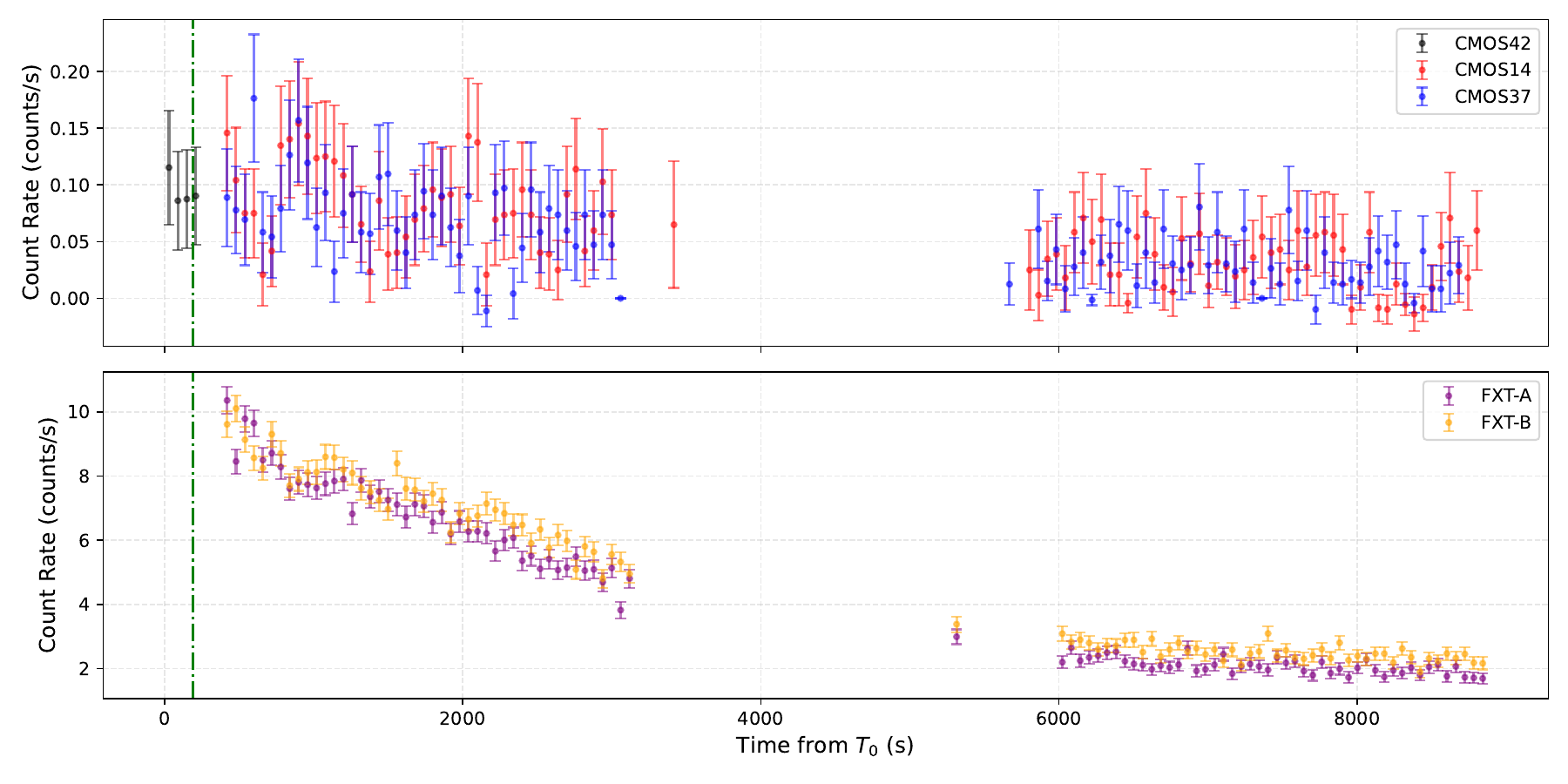}
    \caption{
    \EP/WXT and FXT light curves from
    $T_0=\mathrm{2025\mbox{-}06\mbox{-}05\ 18:48:35.984\ UTC}$
    to $\mathrm{21:17:15.9\ UTC}$ (total span $\simeq 2.3$~h).
    \textit{Top panel:} \EP/WXT light curve in the $0.5$--$4.0$\,keV band.
    Different colours indicate distinct WXT CMOS detectors, as listed in the legend.
    \textit{Bottom panel:} \EP/FXT light curve in the $0.3$--$10.0$\,keV band.
    The purple and orange points indicate data from FXT-A and FXT-B, respectively.
    The two FXT modules are co-aligned.
    The green dash-dotted vertical line in both panels marks the WXT detection/trigger reference time.
    }
    \label{fig:ep_lc}
\end{figure}

The contemporaneous \textit{EP}/FXT observation provides a short X-ray view of RX~J1553.0+4457 after the WXT detection associated with \textit{EP} trigger 01709177873, as shown in the light curves Fig~\ref{fig:ep_lc}.

\begin{figure}[ht]
    \centering
    \includegraphics[width=\columnwidth]{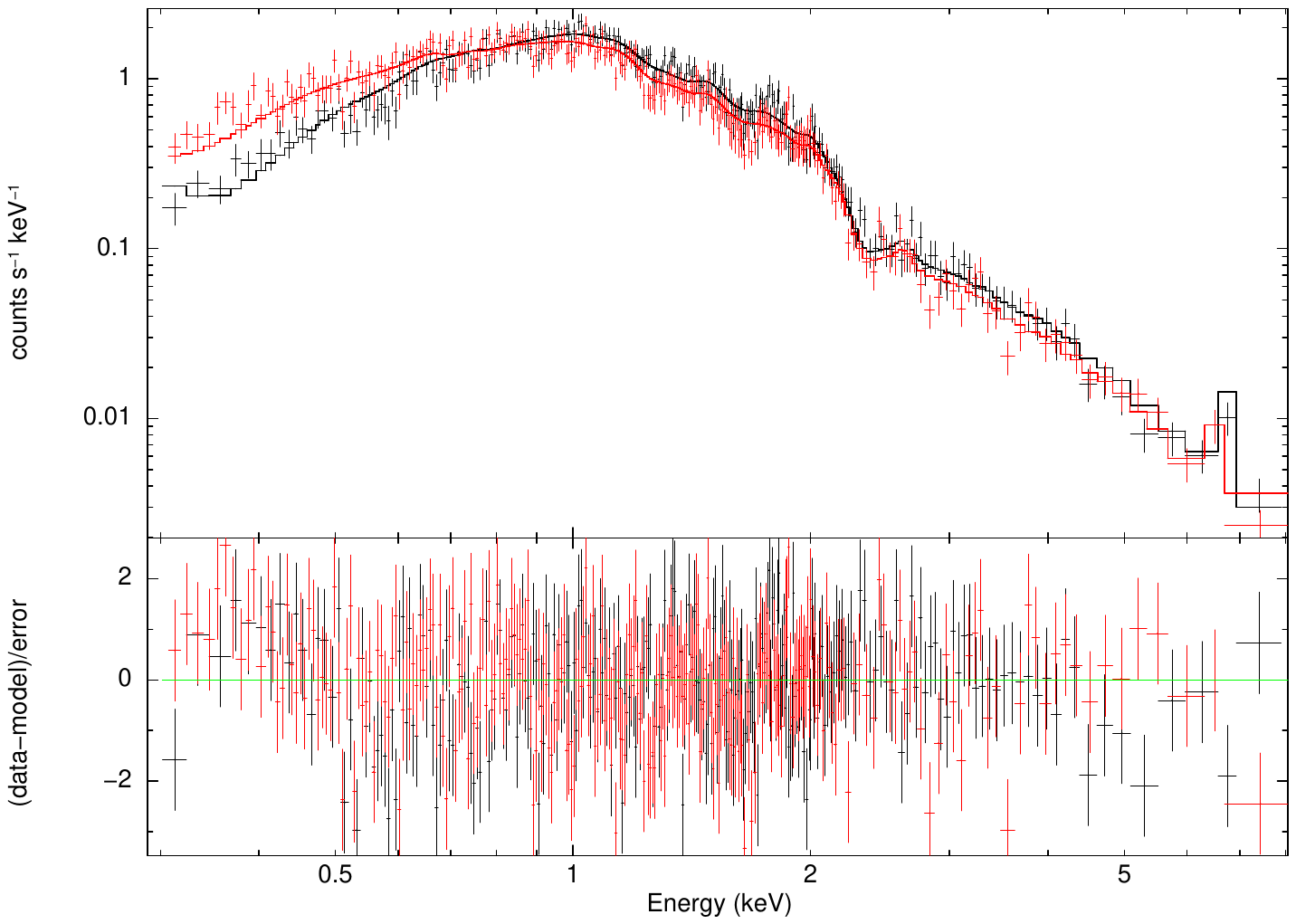}
    \caption{Time-averaged \textit{Einstein Probe}/FXT spectrum of RX~J1553.0+4457 for the full observation, fitted with the three-temperature \textsc{APEC} model adopted in this work. The black and red datasets correspond, respectively, to FXT-A and FXT-B, the two simultaneous, co-pointed FXT modules, and were fitted jointly with linked model parameters. These two spectra therefore represent independent measurements of the same time-averaged observation rather than different source states or time intervals. The upper panel shows the observed \textit{EP}/FXT spectra together with the best-fitting folded models, and the lower panel shows the corresponding fit residuals.}
    \label{fig:ep_fxt_spec}
\end{figure}

\begin{figure}[htbp]
    \centering    
    \includegraphics[width=\columnwidth]{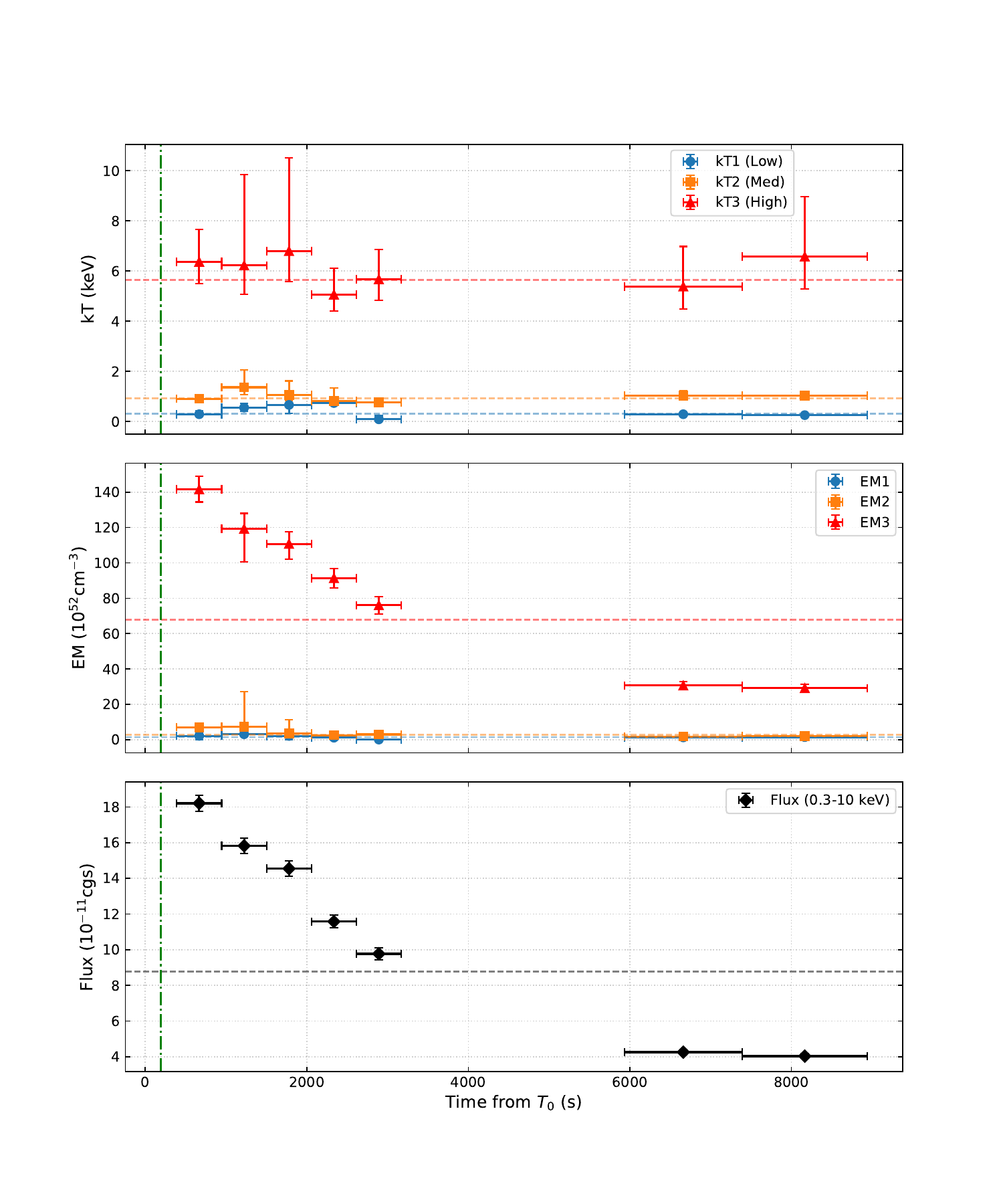}
    \caption{Time-resolved \textit{Einstein Probe}/FXT spectral evolution of RX~J1553.0+4457.
    \textit{Top panel:} Best-fitting temperatures $kT_1$, $kT_2$, and $kT_3$ for the three \textsc{APEC} components in the seven time-resolved \textit{EP}/FXT spectra.
    \textit{Middle panel:} Corresponding emission measures $EM_1$, $EM_2$, and $EM_3$, in units of $10^{52}\,\mathrm{cm^{-3}}$.
    \textit{Bottom panel:} Model-derived $0.3$--$10.0$\,keV flux for each time bin.
    The points are placed at the mid-times of the spectral bins, and the horizontal error bars indicate the widths of the corresponding time intervals.
    The dashed horizontal lines show the best-fitting values from the time-averaged \textit{EP}/FXT spectrum of the full observation.
    The time axis is measured from the start of the \textit{EP}/WXT/CMOS42 exposure,
    $T_0 = \mathrm{2025\mbox{-}06\mbox{-}05\ 18:48:35.984\ UTC}$;
    the green dash-dotted vertical line marks the WXT detection/trigger reference time.}
    \label{fig:ep_fxt_context}
\end{figure}

Figure~\ref{fig:ep_fxt_spec} shows the time-averaged \textit{EP}/FXT spectrum of the full observation together with the best-fitting three-temperature \textsc{APEC} model adopted in this work, while Fig.~\ref{fig:ep_fxt_context} and Table~\ref{tab:ep_specfits} summarize the corresponding time-resolved fits for the seven \textit{EP}/FXT intervals. The time-averaged fit describes the overall spectral shape well, with no significant systematic residuals. The source is brightest in the first resolved interval and then fades substantially over the following bins. The model-derived $0.3$--$10$\,keV flux decreases from $18.2\times10^{-11}$ to $\sim4.0$--$4.3\times10^{-11}\,\mathrm{erg\,cm^{-2}\,s^{-1}}$, corresponding to an overall decline by a factor of about four during the available \textit{EP} coverage. Because the reference time marks the WXT detection/trigger epoch rather than an independently measured X-ray peak, the FXT data show that the source was fading during the available FXT coverage, but they do not by themselves determine the exact flare maximum time. 

Within the uncertainties, the three fitted plasma temperatures remain broadly consistent throughout the observation, with representative values of $\sim0.2$--$0.7$\,keV, $\sim0.8$--$1.4$\,keV, and $\sim5$--$7$\,keV for the low-, intermediate-, and high-temperature components, respectively. The clearest evolution is instead seen in the emission measures, particularly for the hottest component, whose normalization drops from $\sim1.4\times10^{2}$ to $\sim3\times10^{1}\times10^{52}\,\mathrm{cm^{-3}}$ between the first and last resolved intervals. In this phenomenological description, the \textit{EP} data therefore indicate that the short-term X-ray fading is driven mainly by a decrease in the overall emitting plasma, rather than by a strong shift in the characteristic plasma temperatures.

%-----

%-----

%-----

\subsection{Optical emission-line measurements}
\label{subsec:optical_spectroscopy_results}

\begin{table}[htbp]
\centering
\caption{
Single-epoch CAFOS emission-line measurements for RX~J1553.0+4457. Line fluxes are in units of $10^{-14}\,\mathrm{erg\,s^{-1}\,cm^{-2}}$. Equivalent widths are defined as positive for emission. The He\,I~$\lambda7065$ entry is reported as a $3\sigma$ upper limit.
}
\label{tab:cafos_line_measurements}
\small
\setlength{\tabcolsep}{2pt}
\renewcommand{\arraystretch}{1.15}
\begin{tabular}{@{}lcccc@{}}
\hline
Line &
\begin{tabular}{c}Window\\(\AA)\end{tabular} &
Flux &
\begin{tabular}{c}Equivalent\\width\\(\AA)\end{tabular} &
\begin{tabular}{c}Full width\\at half\\maximum\\(\AA)\end{tabular} \\
\hline
H$\gamma$ & 4325.0--4355.0 & $6.65 \pm 0.13$ & $27.23 \pm 0.68$ & $5.8 \pm 0.9$ \\
H$\beta$  & 4845.0--4882.0 & $7.96 \pm 0.24$ & $16.82 \pm 0.66$ & $6.1 \pm 0.9$ \\
H$\alpha$ & 6525.0--6600.0 & $50.30 \pm 4.34$ & $49.21 \pm 4.26$ & $7.4 \pm 0.9$ \\
He\,I~$\lambda7065$ & 7060.0--7072.0 & $<1.94$ & $<1.50$ & -- \\
\hline
\end{tabular}
\end{table}

The CAFOS spectrum shows a red-sloping continuum with Balmer emission superposed on it. Broad molecular absorption structure in the red part of the spectrum is also consistent with a late-type companion contributing strongly to the optical continuum. H$\alpha$ is the strongest detected line, while H$\beta$ and H$\gamma$ are also clearly present in emission. The corresponding single-epoch line fluxes, emission equivalent widths, and full-width-at-half-maximum values are listed in Table~\ref{tab:cafos_line_measurements}. These measurements characterize the Balmer-emitting state at the CAFOS epoch, although the modest spectral resolution and the single-epoch nature of the CAFOS sequence limit a more detailed physical interpretation.

The Balmer-emitting state is qualitatively consistent with the phase-resolved spectroscopy reported by \citet{Liu2025}. Our data, however, consist of a single low-to-moderate-resolution CAFOS sequence and are not suitable for testing the double-peaked line morphology or phase-dependent velocity structure discussed in that work. We therefore use the CAFOS spectrum as near-epoch optical spectroscopic context for the BOOTES and \textit{EP}/FXT activity, and as additional line-state information alongside the long-baseline \textit{TESS} analysis. Around He\,I~$\lambda7065$, the integrated signal is not significant, so we report only a $3\sigma$ upper limit and do not claim a secure He\,I detection. Moreover, because the merged continuum depends on the relative scaling of the sequential B100, G100, and R100 grism segments and represents a composite WD+M-dwarf spectrum, we do not use the CAFOS continuum as an independent constraint on the stellar effective temperatures.

%========================================================%========================================================

\subsection{Spectral energy distribution and two--component fit}
\label{subsec:sed_results}

The adopted two-component \textsc{VOSA} solution provides a satisfactory description of the observed SED from the ultraviolet to the mid-infrared across 18 detected photometric points and the \textit{GALEX} FUV upper limit. In view of the independently established WD+M-dwarf nature of RX~J1553.0+4457 \citep{Liu2025}, we use this fit mainly as a consistency check of the adopted binary configuration and as a test for any additional hot, disc-like, or infrared-excess component. The late-type companion dominates the optical, near-infrared, and mid-infrared fluxes, while the cool WD component mainly affects the ultraviolet and blue end of the model. In the adopted fit, the Koester WD component has $T_{\rm eff}=\wdteff$~K with fixed $\log g=\wdlogg$, while the BT--Settl companion has $T_{\rm eff}=\mdteff$~K and $\log g=\mdlogg$. The binary-fit goodness-of-fit parameter is $V_{\rm gfb}=\vgfb$.

The WD parameters inferred from the broadband SED should be interpreted conservatively. Because the late-type companion contributes most of the optical-to-infrared flux, the SED alone does not provide a strong independent constraint on the WD radius, surface gravity, or temperature. When the WD surface gravity was allowed to vary over a wider range, the formal goodness of fit improved only slightly but the solution moved towards the upper edge of the allowed $\log g$ grid, with trial solutions around $T_{\rm eff}\simeq7250$--$7500$~K and $\log g\simeq8.5$--9.0. We therefore adopt the physically constrained $\log g=\wdlogg$ solution and do not interpret the fitted surface gravity as an independent dynamical mass measurement from broadband photometry alone. In particular, the SED fit is not used to replace the WD mass of $\simeq0.56\,M_\odot$ reported by \citet{Liu2025}. The robust SED result is instead that the observed flux distribution is compatible with a cool WD plus a late-type companion and does not require an additional luminous accretion disc, hot continuum, or infrared-excess component.

For a WD of this approximate mass, such a low temperature corresponds to a relatively cool, evolved WD, with a cooling age expected to be of order Gyr in standard WD cooling calculations \citep[e.g.][]{Fontaine2001,Bedard2020}. This is compatible with a post-common-envelope binary in which the present optical flares and a substantial part of the X-ray emission are driven by magnetic activity on the tidally locked late-type companion, rather than by a luminous accretion flow.

The ultraviolet constraints are also consistent with this cool-WD interpretation. The \textit{GALEX} NUV detection is reproduced by the blue tail of the WD photosphere, with possible minor contribution from chromospheric or flare-related emission from the companion. At the same time, the \textit{GALEX} FUV upper limit disfavours a substantially hotter WD or a bright accretion-powered ultraviolet continuum. Together with the absence of a \textit{WISE} excess, this supports a detached PCEB configuration in which any weak wind-fed or intermittent accretion contribution is not accompanied by a luminous accretion disc or a bright hot continuum component.

Together with the single-epoch CAFOS spectrum, these SED constraints support a detached or weakly interacting configuration. The present data do not require an additional hot continuum, luminous accretion disc, or high-excitation line-emitting component.

\begin{figure}[!t]
  \centering
  \includegraphics[width=0.50\textwidth]{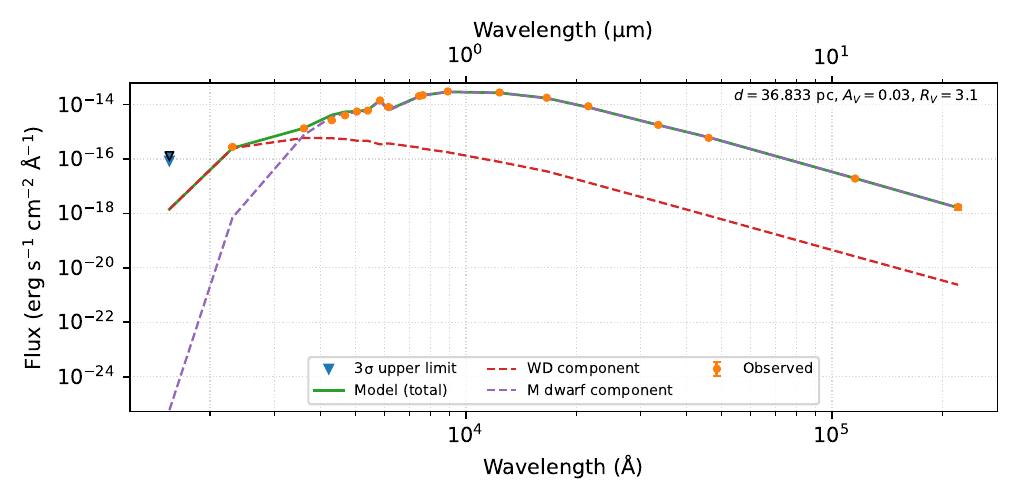}
    \caption{
    Adopted two-component SED fit for RX~J1553.0+4457. The model consists of a
    Koester WD component with $T_{\rm eff}=\wdteff$\,K and fixed
    $\log g=\wdlogg$, and a BT--Settl M-dwarf component with
    $T_{\rm eff}=\mdteff$\,K and $\log g=\mdlogg$. The total model and the two
    stellar components are shown separately; orange points mark detections and
    the blue triangle marks the \textit{GALEX} FUV $3\sigma$ upper limit. The fit adopts
    $d=\distpc$~pc and $A_V=\av$ as fixed input values. The two-component model is used as a consistency check of the known WD+M-dwarf configuration and does not require an additional mid-infrared excess component.
    }
  \label{fig:sed}
\end{figure}

%%%%%%%%%%%%%%%%%%%%%%%%%%%%%%%%%%%%%%%%%%%%%%%%%%%%%%%%%%%%%%%%%%%%%%%%%%%%%%%%%%%%%%%%%%%%%%%%%%%%%%%%%%%%%%%%%%%%%%%%%%%%%%%%%%%%%%%%%%%%%%%%%%%%%%%%%%%%%%%%%%%%%%%%%%%%%%%%%%%%%%%%%%%%%%%%%%%%%%%%%%%%%%%%%%%%%%%%%%%%%%%%%%%%%%%%%%%%%%%%%%%%%%%%%%%%%
\section{Discussion}
\label{sec:discussion}

We interpret our measurements in the context of the established orbital solution and WD+M-dwarf architecture of RX~J1553.0+4457. The new constraints considered here come from diagnostics that are largely independent of that solution: the colour-dependent optical flare decay, the long-baseline \textit{TESS} waveform, the time-resolved X-ray fading, and the absence of any required excess component in the broadband SED.

The BOOTES data provide the most direct evidence that the rapid optical variability can be explained by magnetic activity on the late-type companion. The two short-lived flares show strongly chromatic decay, with systematically faster fading in the bluer bands. This behaviour is expected for impulsive heating followed by cooling in stellar flares and does not, by itself, require an accretion-powered optical origin \citep{Gunther2020,Doyle2019,Kowalski2024}. The band-limited flare energetics and the preference for power-law  over exponential decay are consistent with a stellar-flare origin,  though the single-night BOOTES data do not constrain the long-term  flare duty cycle.

The \textit{TESS} analysis adds a separate long-baseline constraint on the origin of the stable orbital waveform. The dominant signal occurs at $P_{\rm orb}/2$ and remains phase-coherent across six sectors, producing a stable double-wave morphology over a multi-year baseline. This behaviour is naturally expected from ellipsoidal modulation of the tidally locked M-dwarf companion, which fills approximately 93\,\% of its Roche lobe according to the parameters of \citet{Liu2025} ($R_B = 0.403\,R_\odot$, $a = 1.25\,R_\odot$). At this filling factor, tidal distortion produces two photometric minima per orbital cycle, consistent with the observed dominant modulation at $2f_{\rm orb}$. The reported asymmetry between the two minima, corrected in their model by a single cool starspot, does not change this geometric interpretation.

This double-wave interpretation is distinct from irradiation-dominated single-wave orbital variability. Hot-WD PCEBs such as NN~Ser show strong reflection/heating effects from the irradiated face of the companion \citep[e.g.][]{Brinkworth2006,Parsons2010}. Single-wave orbital modulations are also seen in WD pulsars, where they are linked to heating by non-thermal emission from a rapidly spinning, strongly magnetized white dwarf that accelerates electrons impinging on a close M-dwarf companion \citep{Marsh2016,CastroSegura2025}. RX~J1553.0+4457 differs from both cases: its WD is cool ($T_{\rm eff}\simeq7000$\,K), and the observed \textit{TESS} waveform is dominated by the first harmonic rather than by the orbital fundamental. Irradiation from the cool WD is therefore not expected to dominate the optical modulation.

The SED fit and the \textit{EP}/FXT spectral evolution support this interpretation through complementary lines of evidence. The ultraviolet-to-mid-infrared flux distribution is consistent with the independently established cool-WD plus late-type M-dwarf configuration, and the \textit{WISE} points do not require a significant infrared excess. Although the broadband SED alone does not strongly constrain the WD parameters, the \textit{GALEX} FUV upper limit disfavours a substantially hotter WD or a bright accretion-powered ultraviolet continuum. In cataclysmic variables, long-term accretion can heat the WD through compressional heating, making the WD effective temperature a useful tracer of the time-averaged accretion state \citep{TownsleyBildsten2004}. The low WD temperature in RX~J1553.0+4457 is therefore consistent with the absence of sustained high-state accretion and with a system that has had time to cool after the common-envelope phase. At the same time, the \textit{EP}/FXT data show that the system was not X-ray quiet: during the available FXT coverage the model-derived $0.3$--$10$\,keV flux decreased by a factor of about four. In the adopted three-temperature \textsc{APEC} description, this fading is driven mainly by declining emission measures, especially in the hottest component, rather than by a large shift in fitted plasma temperatures. The X-ray behaviour is therefore consistent with fading optically thin plasma, but it does not by itself require a luminous, persistent accretion flow.

Taken together, these constraints favour a near-Roche-lobe-filling pre-cataclysmic binary rather than an actively mass-transferring CV. A weak wind-fed or intermittent accretion contribution can coexist with this detached configuration, as suggested by the Balmer-line behaviour discussed by \citet{Liu2025} and by analogous detached systems with low-level accretion signatures \citep{Tappert2011,Ribeiro2013,Parsons2021}. However, the independent BOOTES, \textit{TESS}, \textit{EP}/FXT, and SED constraints presented here do not require sustained Roche-lobe overflow, a luminous accretion disc, or accretion as the dominant source of the optical variability.

This interpretation also places RX~J1553.0+4457 in a broader context of compact WD+late-type binaries. Some recently identified long-period radio transients, or long-period transients (LPTs), have been associated with cool-WD plus M-dwarf systems, including GLEAM-X~J0704--37 and ILT~J1101+5521 \citep{HurleyWalker2024,Rodriguez2025LPT,Rodriguez2026LPT}. These objects are not direct analogues of RX~J1553.0+4457, because no coherent radio pulses have been reported from this system. The comparison is nevertheless useful: it highlights an emerging population of compact WD+late-type binaries in which cool WDs, magnetic activity, near-Roche-lobe geometries, and possible wind-mediated interactions can coexist. RX~J1553.0+4457 should therefore be viewed as a pre-CV-like system with strong magnetic activity and possible weak wind-fed accretion, not as a confirmed LPT or WD pulsar.

Further progress will require simultaneous phase-resolved spectroscopy and dense optical/X-ray monitoring, potentially including near-infrared diagnostics around He\,I~$\lambda10830$, to separate the roles of magnetic activity, orbital geometry, and any residual accretion \citep[e.g.][]{Fuhrmeister2019He10830}.

%%%%%%%%%%%%%%%%%%%%%%%%%%%%%%%%%%%%%%%%%%%%%%%%%%%%%%%%%%%%%%%%%%%%%%%%%%%%%%%%%%%%%%%%%%%%%%%%%%%%%%%%%%%%%%%%%%%%%%%%%%%%%%%%%%%%%%%%%%%%%%%%%%%%%%%%%%%%%%%%%%%%%%%%%%%%%%%%%%%%%%%%%%%%%%%%%%%%%%%%%%%%%%%%%%%%%%%%%%%%%%%%%%%%%%%%%%%%%%%%%%%%%%%%%%%%%
\section{Conclusions}
\label{sec:conclusions}

We have presented a multi-wavelength study of RX~J1553.0+4457 based on BOOTES high-cadence photometry, public multi-sector \textit{TESS} light curves, contemporaneous \textit{Einstein Probe}/FXT X-ray data, optical spectroscopy, and archival broadband photometry. Adopting the previously established orbital and dynamical picture, our aim was to add independent constraints on the short-timescale optical activity, the long-baseline orbital modulation, the X-ray fading episode, and the broadband SED.

Our main results are as follows.  
(1) The BOOTES light curves reveal two short optical flares with clearly chromatic decay, consistent with rapidly cooling stellar-flare emission. The main event fades more rapidly in the bluer bands, and simple exponential decays are disfavoured relative to power-law-like prescriptions. The corresponding observed-band energies are band-limited and are not interpreted as bolometric flare energies.
(2) The combined \textit{TESS} data show a dominant, locally significant, and stable signal at $P\simeq0.0838$\,d, corresponding to $P_{\rm orb}/2$. The dominance of this first harmonic, together with the stable double-wave folded morphology, indicates that the broadband orbital waveform is dominated by ellipsoidal modulation of the tidally distorted late-type companion rather than by a single-wave irradiation or hot-face component.
(3) The contemporaneous \textit{Einstein Probe}/FXT observation captures a short X-ray fading episode. In the adopted three-temperature \textsc{APEC} description, the model-derived $0.3$--$10$\,keV flux decreases by a factor of about four during the available \textit{EP}/FXT coverage. The fitted plasma temperatures remain broadly consistent within the uncertainties, while the strongest evolution is seen in the emission measures, especially for the hottest component.  
(4) The broadband SED is consistent with the independently known cool-WD plus late-type M-dwarf configuration, with the late-type companion dominating the optical-to-infrared flux. The available photometry does not require a luminous accretion disc, an additional hot continuum component, or a significant mid-infrared excess.
(5) The CAFOS spectrum provides single-epoch measurements of the Balmer-emitting state. H$\alpha$, H$\beta$, and H$\gamma$ are detected in emission, with H$\alpha$ showing the largest line flux and equivalent width. We do not find a secure He\,I~$\lambda7065$ detection and report only an upper limit for that feature.

Overall, the data favour a near-Roche-lobe-filling pre-cataclysmic binary in which the stable broadband orbital waveform is dominated by ellipsoidal modulation of the late-type companion, while the short-timescale optical flares are consistent with magnetic activity on that star. The \textit{EP}/FXT detection shows that the system was X-ray active during the available FXT coverage, and a weak wind-fed or intermittent accretion contribution cannot be excluded. However, the cool WD, the SED constraints, and the harmonic structure of the \textit{TESS} waveform do not require sustained Roche-lobe overflow, a luminous accretion disc, or accretion as the dominant explanation for the optical behaviour.

Disentangling the respective contributions of magnetic activity and wind-fed accretion will require simultaneous phase-resolved spectroscopy combined with coordinated X-ray and optical monitoring.

%%%%%%%%%%%%%%%%%%%%%%%%%%%%%%%%%%%%%%%%%%%%%%%%%%%%%
\section*{Data availability}
The \textit{TESS} full-frame image cutouts analysed in this work are publicly available from the Mikulski Archive for Space Telescopes (MAST; \citealt{Marston2018}) via \texttt{TESScut} \citep{Brasseur2019}. The archival photometry compiled for the SED analysis comes from \textit{GALEX}, SDSS, APASS, \textit{Gaia} DR3, 2MASS, and \textit{WISE}/\textit{NEOWISE} \citep{Martin2005,morrissey2007,Alam2015,Henden2016,GaiaCollaboration2023,Skrutskie2006,wright2010,Mainzer2011}. The BOOTES photometric data, the reduced optical spectrum, the processed \textit{Einstein Probe}/FXT products used in this study, and the derived data products underlying the figures and tables are available from the corresponding author upon reasonable request.

\begin{acknowledgements}
We acknowledge the use of data from the BOOTES network and thank the Instituto de Astrof\'{\i}sica de Andaluc\'{\i}a (IAA-CSIC) for its support and collaboration. The BOOTES-4/MET telescope is located at Lijiang Astronomical Observatory, the BOOTES-5/JGT telescope at the Observatorio Astron\'omico Nacional in San Pedro M\'artir, the BOOTES-6/DPRT telescope at Boyden Observatory, and the BOOTES-7 telescope at San Pedro de Atacama Observatory; we thank the staff of these observatories for their support. This work is based in part on observations collected at the Centro Astron\'omico Hispano en Andaluc\'{\i}a at Calar Alto, proposal 25A-2.2-013, operated jointly by Junta de Andaluc\'{\i}a and Consejo Superior de Investigaciones Cient\'{\i}ficas (IAA-CSIC).

This work is based in part on data obtained with the \textit{Einstein Probe} mission, a mission of the Chinese Academy of Sciences supported by the Strategic Priority Program on Space Science of the Chinese Academy of Sciences in collaboration with ESA, MPE, and CNES. We acknowledge the data resources and technical support provided by the China National Astronomical Data Center, the Astronomical Data Center of the Chinese Academy of Sciences, and the Chinese Virtual Observatory.

This paper includes data from the \textit{TESS} mission and the \textit{Galaxy Evolution Explorer}, obtained from the Mikulski Archive for Space Telescopes (MAST) at the Space Telescope Science Institute. Funding for the \textit{TESS} mission is provided by the NASA Explorer Program. This research also made use of data products from the Zwicky Transient Facility, obtained through the NASA/IPAC Infrared Science Archive, and of data from the ESA mission \textit{Gaia}, processed by the \textit{Gaia} Data Processing and Analysis Consortium. This publication also makes use of data products from the Two Micron All Sky Survey and the \textit{Wide-field Infrared Survey Explorer}.

S.-Y. Wu acknowledges financial support from the China Scholarship Council (CSC). AJCT acknowledges the Spanish Ministry of Science, Innovation and Universities project PID2023-151905OB-I00. MG thanks Adolfo González Rivera (Alhama Academy) for logistical support and acknowledges funding from the Academy of Finland (project no. 325806). The programme of development within Priority-2030 is acknowledged for supporting the research at UrFU. MCG acknowledges financial support from the Spanish Ministry project MCI/AEI/PID2023-149817OB-C31 and the Severo Ochoa grant CEX2021-001131-S funded by MICIU/AEI/10.13039/501100011033. NCS acknowledges support from the Science and Technology Facilities Council grant ST/X001121/1. LHG acknowledges financial support from ANID program FONDECYT Iniciaci\'{o}n 11241477.
\end{acknowledgements}

%%%%%%%%%%%%%%%%%%%%
\bibliographystyle{aa}   
\bibliography{references}

%%%%%%%%%%%%%%%%%%%%%%%%%%%%%%%%%%%%%
\onecolumn
\begin{appendix}
\section{Photometric points used in the SED fit}
\label{app:phot}

Table~\ref{tab:sed_used_main} lists the photometric measurements used in our SED fit (19 entries in total), including 18 detections and one $3\sigma$ upper limit. The table reports the catalogue fluxes before dereddening, in $\mathrm{erg\,s^{-1}\,cm^{-2}}\,\text{\AA}^{-1}$; in the \textsc{VOSA} SED analysis these fluxes were dereddened using the fixed input value $A_V=\av$. Upper limits are indicated by the ``$<$'' symbol and their uncertainties are omitted. ``Source'' refers to the originating survey/filter family. The table includes the \textit{GALEX} far-ultraviolet (FUV) and near-ultraviolet (NUV) bands, optical and near-infrared survey bands, and \textit{WISE} mid-infrared bands. All listed entries are included in the final SED fit; therefore, no flag column is provided. The archival photometry compiled for the SED analysis comes from \textit{GALEX}, SDSS, APASS, \textit{Gaia} DR3, 2MASS, and \textit{WISE}/\textit{NEOWISE} \citep{Martin2005,morrissey2007,Alam2015,Henden2016,GaiaCollaboration2023,Skrutskie2006,wright2010,Mainzer2011}.

\begin{table}[htbp]
\centering
\caption{Photometric points used in the SED fit (19 entries; 18 detections and one upper limit). The listed values are catalogue fluxes before dereddening, in $\mathrm{erg\,s^{-1}\,cm^{-2}}\,\text{\AA}^{-1}$. In the \textsc{VOSA} fit these points were dereddened using the fixed input value $A_V=\av$.}
\label{tab:sed_used_main}

\setlength{\tabcolsep}{4pt}
\renewcommand{\arraystretch}{1.4}
\begin{tabular}{lcccc}
\hline

Filter & $\lambda_{\rm eff}$ [\AA] & Flux & $\Delta$Flux & Source \\

\hline

\textit{GALEX} $FUV$ & 1549 & $<8.49\times 10^{-17}$ & -- & \textit{GALEX} \\
\textit{GALEX} $NUV$ & 2303 & $2.77\times 10^{-16}$ & $2.32\times 10^{-17}$ & \textit{GALEX} \\
SDSS $u$ & 3608 & $1.32\times 10^{-15}$ & $1.10\times 10^{-17}$ & Sloan \\
APASS $B$ & 4299 & $2.63\times 10^{-15}$ & $1.36\times 10^{-16}$ & Misc \\
SDSS $g$ & 4672 & $3.98\times 10^{-15}$ & $2.19\times 10^{-16}$ & Sloan \\
\textit{Gaia} $G_{BP}$ & 5036 & $5.50\times 10^{-15}$ & $5.43\times 10^{-17}$ & \textit{Gaia} \\
APASS $V$ & 5394 & $5.94\times 10^{-15}$ & $1.64\times 10^{-16}$ & Misc \\
\textit{Gaia} $G$ & 5822 & $1.41\times 10^{-14}$ & $4.66\times 10^{-17}$ & \textit{Gaia} \\
SDSS $r$ & 6141 & $8.08\times 10^{-15}$ & $4.05\times 10^{-16}$ & Sloan \\
SDSS $i$ & 7458 & $2.05\times 10^{-14}$ & $1.31\times 10^{-15}$ & Sloan \\
\textit{Gaia} $G_{RP}$ & 7620 & $2.20\times 10^{-14}$ & $1.62\times 10^{-16}$ & \textit{Gaia} \\
SDSS $z$ & 8923 & $3.01\times 10^{-14}$ & $8.61\times 10^{-17}$ & Sloan \\
2MASS $J$ & 12350 & $2.75\times 10^{-14}$ & $5.32\times 10^{-16}$ & 2Mass \\
2MASS $H$ & 16620 & $1.76\times 10^{-14}$ & $2.76\times 10^{-16}$ & 2Mass \\
2MASS $Ks$ & 21590 & $8.58\times 10^{-15}$ & $1.58\times 10^{-16}$ & 2Mass \\
\textit{WISE} $W1$ & 33526 & $1.78\times 10^{-15}$ & $3.31\times 10^{-17}$ & \textit{WISE} \\
\textit{WISE} $W2$ & 46028 & $5.93\times 10^{-16}$ & $9.60\times 10^{-18}$ & \textit{WISE} \\
\textit{WISE} $W3$ & 115608 & $1.91\times 10^{-17}$ & $3.86\times 10^{-19}$ & \textit{WISE} \\
\textit{WISE} $W4$ & 220883 & $1.72\times 10^{-18}$ & $3.69\times 10^{-19}$ & \textit{WISE} \\

\hline
\end{tabular}
\end{table}

\section{Completeness and robustness checks}
\label{app:completeness}

To quantify the selection function of the \textit{TESS} flare sample, we performed injection--recovery tests on the detrended residual light curves after removal of the dominant orbital waveform. For each injected flare, we recorded both the raw detection outcome and the final clean-sample recovery outcome. The completeness threshold adopted in the main text is defined from the clean-recovery curve rather than from the raw-detection curve, because the final flare sample is subject to additional selection cuts in peak amplitude, signal-to-noise ratio, duration, and equivalent duration.

The resulting clean-sample completeness curve yields ${\rm ED}_{50}\approx21.7$~s and ${\rm ED}_{80}\approx40.4$~s, while ${\rm ED}_{90}$ is not reached within the explored ED grid. Of the 4320 attempted injections, 4228 are recovered at the raw-detection stage and 2552 remain in the final clean sample. The limiting factor is therefore not raw detectability alone, but also the conservative event-definition criteria applied to the final sample.

The cumulative flare-frequency distribution (FFD) was constructed from the clean flares above the corresponding ${\rm ED}_{80}$ completeness threshold. For each duration-cut sample, we modelled it using Eq.~\eqref{eq:ffd_powerlaw},
\begin{equation}
    N(>{\rm ED}) \propto {\rm ED}^{\beta},
    \label{eq:ffd_powerlaw}
\end{equation}
where $N(>{\rm ED})$ is the cumulative number of flares with equivalent duration larger than ${\rm ED}$. The direct slope $\beta_{\rm direct}$ was obtained from a linear regression in $\log N(>{\rm ED})$ versus $\log {\rm ED}$ above ${\rm ED}_{80}$. The bootstrap slope $\beta_{\rm boot}$ was obtained by resampling the clean flare list and repeating the same fit; the quoted intervals give the 16th--84th percentile range.

As an additional robustness test, we repeated the clean-sample and FFD analysis for several duration-cut choices. Although the inferred completeness threshold changes moderately, the best-fitting cumulative FFD slope remains effectively unchanged within the tested range, indicating that the main ED-space FFD result is not sensitive to the exact duration cut adopted.

\begin{figure*}[htbp]
\centering
\includegraphics[width=0.82\textwidth]{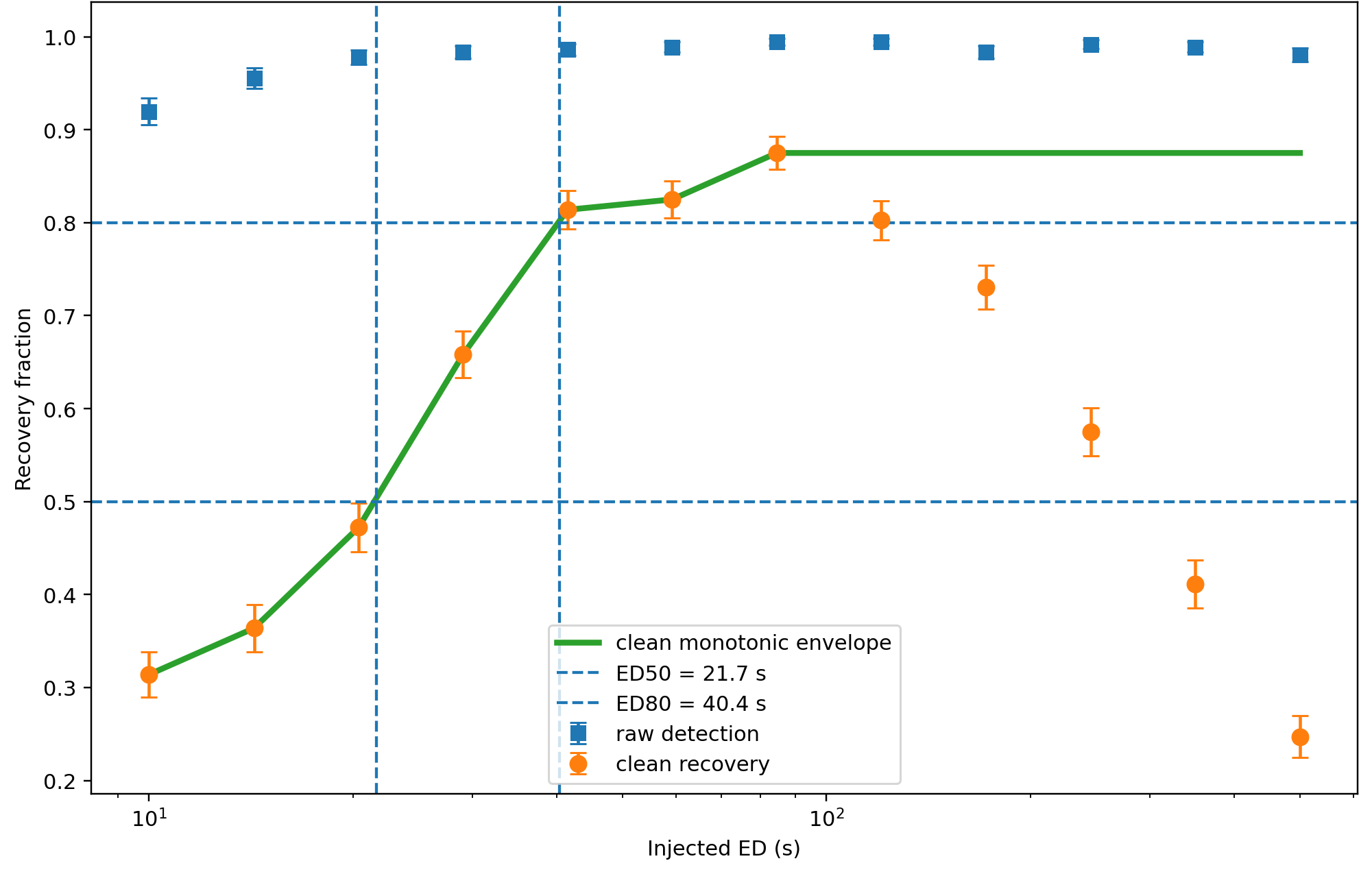}
\caption{Injection--recovery completeness for the \textit{TESS} flare search of RX~J1553.0+4457. Blue squares show the raw detection fraction, while orange circles show the clean-recovery fraction after the final sample-definition cuts. The green curve marks the monotonic envelope adopted for the clean-sample completeness estimate. The dashed lines indicate the corresponding ${\rm ED}_{50}$ and ${\rm ED}_{80}$ thresholds derived from the clean-recovery envelope. The figure shows that raw detection is efficient over most of the explored ED range, whereas the completeness of the final clean sample is additionally limited by the conservative event-selection criteria.}
\label{fig:appendix_completeness}
\end{figure*}

\begin{table*}[htbp]
\centering
\scriptsize
\caption{\textit{TESS} clean flare catalogue for RX~J1553.0+4457. The equivalent duration is measured from the orbital-model-subtracted relative-excess light curve. $E_{\rm TESS}$ is the flare energy in the \textit{TESS} band, while $E_{\rm bol,app}$ is the approximate bolometric energy obtained using $E_{\rm bol,app}=4.8\,E_{\rm TESS}$, following the bandpass correction adopted by \citet{Vida2019}, for comparison with the literature flare sample.}
\label{tab:tess_flare_catalogue}

\setlength{\tabcolsep}{4pt}
\renewcommand{\arraystretch}{1.15}
\begin{tabular}{lcccccccc}
\hline
Flare ID & Sector & $t_{\rm peak}$ & Duration & Peak amp. & Peak S/N & ED & $E_{\rm TESS}$ & $E_{\rm bol,app}$ \\
 &  & (BJD$-2457000$) & (min) & (\%) &  & (s) & ($10^{33}$ erg) & ($10^{34}$ erg) \\
\hline
S23\_M001 & 23 & 1928.18011 & 149.9 & 4.61 & 18.8 & 164.2 & 3.53 & 1.70 \\
S23\_M002 & 23 & 1945.99286 & 120.0 & 4.15 & 16.9 & 129.0 & 2.78 & 1.33 \\
S23\_M003 & 23 & 1954.15955 & 120.0 & 10.77 & 43.8 & 252.1 & 5.43 & 2.61 \\
S24\_M001 & 24 & 1973.88800 & 59.9 & 2.71 & 9.8 & 48.9 & 1.06 & 0.50 \\
S50\_M001 & 50 & 2671.31182 & 110.0 & 4.74 & 14.4 & 114.4 & 2.47 & 1.18 \\
S50\_M002 & 50 & 2674.33965 & 30.0 & 4.52 & 13.7 & 30.5 & 0.65 & 0.32 \\
S50\_M003 & 50 & 2680.67306 & 140.0 & 3.21 & 9.7 & 129.6 & 2.79 & 1.34 \\
S50\_M004 & 50 & 2682.93002 & 130.0 & 10.08 & 30.6 & 173.7 & 3.74 & 1.79 \\
S51\_M001 & 51 & 2700.84670 & 50.0 & 2.96 & 8.4 & 41.6 & 0.90 & 0.43 \\
S51\_M002 & 51 & 2700.96476 & 50.0 & 5.05 & 14.4 & 51.0 & 1.10 & 0.53 \\
S77\_M001 & 77 & 3399.25542 & 53.3 & 15.20 & 42.9 & 77.7 & 1.67 & 0.80 \\
S77\_M002 & 77 & 3404.05642 & 40.0 & 5.13 & 14.5 & 30.6 & 0.65 & 0.32 \\
S78\_M001 & 78 & 3434.71025 & 46.7 & 5.58 & 16.1 & 59.0 & 1.27 & 0.61 \\
\hline
\end{tabular}
\tablefoot{
The final clean sample requires peak S/N $>8$, peak amplitude $>2\%$, duration $<180$~min, and ${\rm ED}>30$~s. Candidates separated by less than 60~min are merged before the final event measurement. The listed \textit{TESS}-band energies use the same quiescent luminosity adopted for the flare-frequency analysis; the approximate bolometric energies are used only for comparison with the \citet{Gunther2020} flare-energy distribution.}
\end{table*}

\begin{table*}
\caption{Sensitivity of the \textit{TESS} flare completeness and cumulative FFD slope to the adopted maximum-duration cut. The reference clean sample uses a maximum duration of 180 min. The slope $\beta$ is measured from the cumulative relation $N(>{\rm ED})\propto {\rm ED}^{\beta}$ above the corresponding ${\rm ED}_{80}$ threshold.}
\label{tab:tess_duration_sensitivity}
\centering
\begin{tabular}{lcccccc}
\hline
Duration cut & $N_{\rm flare}$ & ${\rm ED}_{50}$ (s) & ${\rm ED}_{80}$ (s) & ${\rm ED}_{90}$ (s) & $\beta_{\rm direct}$ & $\beta_{\rm boot}$ \\
\hline
180 min
& 13
& 21.50
& 41.05
& n.r.
& $-1.16$
& $-1.18^{+0.15}_{-0.18}$ \\
240 min
& 13
& 20.40
& 39.02
& 76.07
& $-1.16$
& $-1.19^{+0.15}_{-0.17}$ \\
300 min
& 13
& 19.74
& 37.60
& 70.33
& $-1.16$
& $-1.17^{+0.14}_{-0.18}$ \\
No duration cut
& 13
& 17.99
& 35.55
& 61.07
& $-1.16$
& $-1.20^{+0.15}_{-0.19}$ \\
\hline
\end{tabular}
\tablefoot{``n.r.'' means that the 90 per cent clean-recovery threshold is not reached within the explored equivalent-duration grid.}
\end{table*}

%%%%%%%%%%%%%%%%%%%%%%%%%%%%%%%%%%%%%%%%%%%%

\begin{table*}[t]
\centering
\scriptsize
\caption{Best-fitting parameters of the time-averaged and time-resolved \textit{EP}/FXT spectra of RX~J1553.0+4457. Emission measures are given in units of $10^{52}\,\mathrm{cm^{-3}}$.}
\label{tab:ep_specfits}

\setlength{\tabcolsep}{3pt}
\renewcommand{\arraystretch}{1.5}

\begin{tabular}{lcccccccc}
\hline
Time from $T_2$ & $kT_1$ & $EM_1$ & $kT_2$ & $EM_2 $ & $kT_3$ & $ EM_3$ & C-stat/d.o.f. & Flux$_{0.3-10.0\mathrm{keV}}$ \\
\multicolumn{1}{c}{(s)} & (keV) & ($10^{52}\,\mathrm{cm^{-3}}$) & (keV) & ($10^{52}\,\mathrm{cm^{-3}}$) & (keV) & ($10^{52}\,\mathrm{cm^{-3}}$) & &($10^{-11} \mathrm{erg\cdot cm^{-2}\cdot s^{-1}}$)\\
\hline
$0-8530.3$ &
$0.310^{+0.072}_{-0.062}$ &
$1.403^{+0.464}_{-0.545}$ &
$0.922^{+0.065}_{-0.063}$ &
$2.772^{+0.349}_{-0.350}$ &
$5.643^{+0.382}_{-0.335}$ &
$67.858^{+1.547}_{-1.503}$ &
$578.14/539\,(1.07)$ & $8.776_{-0.102}^{+0.094}$\\

$0-556$&
$0.290^{+0.128}_{-0.090}$ &
$2.170^{+2.090}_{-2.162}$ &
$0.907^{+0.089}_{-0.081}$ &
$7.046^{+1.507}_{-1.343}$ &
$6.367^{+1.289}_{-0.871}$ &
$141.530^{+7.396}_{-7.107}$ &
$341.31/304\,(1.12)$ & $18.203_{-0.457}^{+0.464}$\\

$556-1112$ &
$0.555^{+0.168}_{-0.158}$ &
$3.082^{+1.109}_{-1.113}$ &
$1.364^{+0.682}_{-0.293}$ &
$7.263^{+19.80}_{-4.731}$ &
$6.225^{+3.628}_{-1.150}$ &
$119.233^{+8.735}_{-18.641}$ &
$256.06/274\,(0.93)$ & $15.820_{-0.418}^{+0.435}$\\

$1112-1668$&
$0.662^{+0.267}_{-0.336}$ &
$2.118^{+1.605}_{-2.118}$ &
$1.054^{+0.564}_{-0.231}$ &
$3.475^{+7.968}_{-1.716}$ &
$6.798^{+3.725}_{-1.221}$ &
$110.746^{+6.716}_{-8.803}$ &
$245.45/248\,(0.99)$ & $14.541_{-0.410}^{+0.426}$\\

$1668-2224$ &
$0.737^{+0.087}_{-0.097}$ &
$1.142^{+2.064}_{-1.142}$ &
$0.817^{+0.520}_{-0.175}$ &
$2.553^{+1.428}_{-1.633}$ &
$5.060^{+1.056}_{-0.654}$ &
$91.341^{+5.543}_{-5.452}$ &
$228.01/217\,(1.05)$ & $11.575_{-0.351}^{+0.360}$\\

$2224-2773.2$ &
$0.188^{+0.142}_{-0.069}$ &
$0.074^{+1.716}_{-0.074}$ &
$0.758^{+0.127}_{-0.098}$ &
$2.977^{+0.592}_{-0.958}$ &
$5.670^{+1.185}_{-0.828}$ &
$76.061^{+4.921}_{-4.855}$ &
$187.47/185\,(1.01)$ & $9.768_{-0.331}^{+0.339}$\\

$5557.2-7000$&
$0.290^{+0.085}_{-0.061}$ &
$1.301^{+0.580}_{-0.617}$ &
$1.032^{+0.187}_{-0.150}$ &
$1.768^{+1.054}_{-0.585}$ &
$5.381^{+1.599}_{-0.885}$ &
$30.618^{+2.156}_{-2.164}$ &
$226.56/214\,(1.06)$ & $4.255_{-0.131}^{+0.134}$\\

$7000-8530.3$&
$0.254^{+0.061}_{-0.066}$ &
$1.476^{+0.566}_{-0.557}$ &
$1.028^{+0.137}_{-0.116}$ &
$2.049^{+0.773}_{-0.528}$ &
$6.576^{+2.387}_{-1.291}$ &
$29.132^{+2.090}_{-2.048}$ &
$181.20/202\,(0.89)$ & $4.033_{-0.126}^{+0.129}$\\
\hline
\end{tabular}
\tablefoot{
Times are measured from the start of the \textit{EP}/FXT observation, $T_2=\mathrm{2025\mbox{-}06\mbox{-}05\ 18:55:05.61\ UTC}$. The WXT detection/trigger reference time is $T_1=\mathrm{2025\mbox{-}06\mbox{-}05\ 18:51:47\ UTC}$. This reference time should not be interpreted as an independently measured X-ray flare maximum. The first row gives the fit to the time-averaged \textit{EP}/FXT spectrum, while Bins~1--7 correspond to the seven time-resolved spectra shown in Fig.~\ref{fig:ep_fxt_context}. The numbers in parentheses after C-stat/d.o.f.\ give the corresponding reduced statistic for visual reference only. All fits used a fixed Galactic column density of $N_{\rm H}=8.83\times10^{18}\,\mathrm{cm^{-2}}$; $N_{\rm H}$ was not a fitted parameter and is therefore not listed as a table column. The quoted uncertainties are statistical intervals for the adopted three-temperature \textsc{APEC} model, equivalent to the 90\% confidence region for the parameters. Some emission-measure uncertainties, most notably EM$_2$ in Bin~2 and the hottest component in several time bins, are strongly asymmetric. This behaviour mainly reflects covariance among the temperatures and normalizations of neighbouring thermal components: a weakly constrained hot tail can be partly compensated by changes in its emission measure or by the normalization of an adjacent component. These asymmetric intervals should therefore be interpreted as model-parameter degeneracies within the adopted phenomenological description, rather than as evidence that the X-ray fading trend or the integrated flux measurements are unreliable. For this reason, the physical interpretation in the text focuses on the integrated $0.3$--$10$\,keV flux evolution and the overall spectral behaviour, rather than on small bin-to-bin changes in individual \textsc{APEC} emission measures.}
\end{table*}

\end{appendix}

\end{document}